\documentclass[aps, prl, nobibnotes, superscriptaddress, onecolumn, amsmath, notitlepage]{revtex4-1} 

\usepackage{graphicx}
\usepackage{dcolumn}
\usepackage{bm}
\usepackage{xr-hyper}
\usepackage[hidelinks]{hyperref}
\usepackage{natbib}
\usepackage{color}
\usepackage{amsmath}
\usepackage{amssymb}
\usepackage{xfrac} 
\usepackage{xfp} 
\usepackage{titletoc}

\begin{document}

\title{Emergent resonances in a thin film tailored by optically-induced small permittivity asymmetries} 

\author{Rodrigo Bert\'{e}} \altaffiliation{Contributed equally to this work}
\affiliation	{Chair in Hybrid Nanosystems, Nanoinstitut M\"{u}nchen, Fakult\"{a}t f\"{u}r Physik, Ludwig-Maximilians-Universit\"{a}t M\"{u}nchen, 80799 M\"{u}nchen, Germany}

\author{Thomas Possmayer} \altaffiliation{Contributed equally to this work}
\affiliation	{Chair in Hybrid Nanosystems, Nanoinstitut M\"{u}nchen, Fakult\"{a}t f\"{u}r Physik, Ludwig-Maximilians-Universit\"{a}t M\"{u}nchen, 80799 M\"{u}nchen, Germany}

\author{Andreas Tittl} 
\affiliation	{Chair in Hybrid Nanosystems, Nanoinstitut M\"{u}nchen, Fakult\"{a}t f\"{u}r Physik, Ludwig-Maximilians-Universit\"{a}t M\"{u}nchen, 80799 M\"{u}nchen, Germany}

\author{Leonardo de S. Menezes}
\affiliation	{Chair in Hybrid Nanosystems, Nanoinstitut M\"{u}nchen, Fakult\"{a}t f\"{u}r Physik, Ludwig-Maximilians-Universit\"{a}t M\"{u}nchen, 80799 M\"{u}nchen, Germany}
\affiliation	{Departamento de F\'{i}sica, Universidade Federal de Pernambuco, Recife, Pernambuco 50670-901, Brazil}

\author{Stefan A. Maier}
\affiliation	{School of Physics and Astronomy, Monash University, Clayton, Victoria 3800, Australia}
\affiliation	{The Blackett Laboratory, Department of Physics, Imperial College London, London SW7 2AZ, United Kingdom}

{
\let\clearpage\relax
\maketitle
}

\linespread{1.2}	

\section{Abstract}

Resonances are usually associated with finite systems - the vibrations of clamped strings in a guitar or the optical modes in a cavity defined by mirrors. In optics, resonances may be induced in infinite continuous media via periodic modulations of their optical properties. Here we demonstrate that periodic modulations of the permittivity of a featureless thin film can also act as a symmetry breaking mechanism, allowing the excitation of photonic \mbox{\textit{quasi}-bound} states in the continuum (\textit{q}BICs). By interfering two ultrashort laser pulses in the unbounded film, transient resonances can be tailored through different parameters of the pump beams. We show that the system offers resonances tunable in wavelength and quality-factor, and spectrally selective enhancement of third harmonic generation. Due to a fast decay of the permittivity asymmetry, we observe ultrafast dynamics, enabling time-selective near-field enhancement with picosecond precision. Optically-induced permittivity asymmetries may be exploited in on-demand weak to ultrastrong light-matter interaction regimes and light manipulation at dynamically chosen wavelengths in \mbox{lithography-free metasurfaces}.

\newpage

\section{Introduction}

While individual atoms display spectral lines \cite{CondonBook1951}, electronic and phononic interactions in extended matter generate broad optical responses \cite{Fox2002}. Bulk excitonic resonances are a prominent exception, but their wavelength is predefined near the band edges of the given material \cite{Kazi2014}, severely limiting the optical bandwidth available for applications. A new paradigm in optical resonances appeared with the ability to pattern materials below or in the same size scale as the wavelength of light \cite{NovotnyBook}, into flat (metasurfaces \cite{Kildishev2013}) and into three-dimensional (metamaterials \cite{Kadic2019}) structures, in which optical functionalities are not limited by the weakly-dispersive macroscopic permittivity of its \mbox{constituents. However,} their optical response is intimately linked to the dimensions of the comprising resonators established during fabrication.

Optical modes in infinite featureless media have been shown to arise from periodic modulations of the permittivity of thin films in the seminal theoretical works on guided-mode resonances (GMRs) \cite{Wang1990, Magnusson1992, Wang1993}, in which narrower linewidths were observed for vanishing permittivity modulations ($\Delta\varepsilon \rightarrow 0$). Periodic modulations, recently explored in time varying metamaterials \cite{Galiffi2022}, have also been extensively investigated using laser beam interference in dynamic gratings induced in featureless films \cite{EichlerBook}. Such induced gratings are commonly measured via diffraction effects that are almost invariant for small changes in the wavelength of the probing beam. Here we demonstrate that periodic permittivity changes act as a symmetry breaking mechanism, enabling all-optical tailored resonances. These correspond to \mbox{symmetry-protected permittivity-asymmetric \textit{quasi}-bound states in the continuum ({$\varepsilon$}-\textit{q}BICs)}, for which higher quality-factors ($Q$) correlate with smaller permittivity asymmetries \cite{Liu2020, Yu2022, Berte2023, Li2023}, akin to GMRs \cite{Wang1990, Magnusson1992, Wang1993}. But while GMRs are excited in hybrid grating/waveguide structures with predefined periodicities \cite{Quaranta2018}, we demonstrate a higher control over {$\varepsilon$}-\textit{q}BICs (in wavelength, $Q$ and amplitude) in an all-optical manner without any \mbox{in-plane geometrical constraints.}

Arising in the context of electronic wavefunction localization \cite{NeumannWigner1929}, BICs are a set of phenomena demonstrated in elastic \cite{Lim1969}, acoustic \cite{Parker1966}, hydrodynamic \cite{Ursell1951} and optical \cite{Plotnik2011} systems, where particular oscillatory excitations remain spatially localized in spite of co-existing in the continuum of possible propagating modes. True BICs can only exist in systems devoid of intrinsic losses and that are spatially infinite in at least one dimension \cite{Hsu2016}. Among the different types of BICs \cite{Hsu2016} are those whose symmetry is incompatible with that of propagating modes \cite{Koshelev2018}, thus termed symmetry-protected BICs. As true BICs cannot be measured, their observable counterpart in finite photonic systems, called \textit{quasi}-BICs (\textit{q}BICs) \cite{Rybin2017}, are usually probed via geometrical perturbations in the unit cell of metasurfaces \cite{Fedotov2007, Koshelev2018, Kuhner2022, Noda2001, Hirose2014, Liu2019}.  Recently, \textit{q}BICs have been proposed and theoretically investigated \cite{Liu2020, Yu2022, Berte2023, Li2023} and experimentally demonstrated \cite{Berte2023} in metasurfaces that are geometrically symmetric under an in-plane inversion, but permittivity asymmetric under the same spatial transformation ($C_2$ or $(x,y) \rightarrow (-x,-y)$).

In a similar fashion to dynamic gratings, we are able to induce a permittivity symmetry breaking in an in-plane unbounded homogeneous medium by interfering two $\approx$ 180 fs ultrashort laser pulses on a thin film (Fig.~\ref{fig1}, and Fig.~\ref{figs_optical_setup} of the Supplementary Information - SI - for the optical setup used in the pump-probe experiments). 

\newpage
\begin{figure}[h] 
	\centering
	\includegraphics[scale=0.97]{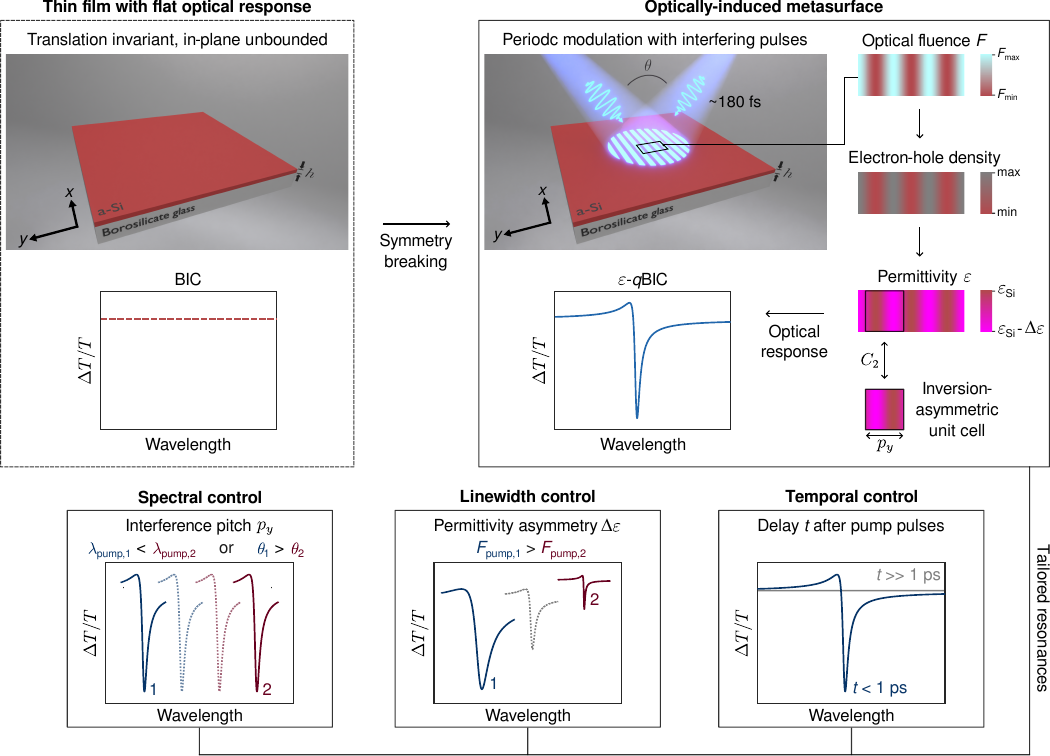}
	\caption{\textbf{Permittivity symmetry breaking of continuous media and tailored $\varepsilon$-\textit{q}BICs.} (top left) An unbounded thin film of continuous translational symmetry features a symmetry-protected BIC state, which does not couple to a far-field impinging radiation (flat optical response). (top right) Scheme of the symmetry breaking through the interference of two \mbox{$\approx$ 180 fs} pulses on the a-Si thin film (of height $h$) deposited on a borosilicate cover glass. The pump beams interfere at an angle $\theta$ generating a 1D grating pattern of bright (BF) and dark fringes (DF) that excites electron-hole pairs ($e^-/h^+$) mainly at the BFs via absorption (middle panel). These excited carriers impose a periodicity $p_y$ in the film permittivity (bottom panel), whose 1D unit cell is asymmetric under an in-plane inversion ($C_2$ or $(x,y) \rightarrow (-x,-y)$) transformation. The optically-induced metasurface is able to temporarily sustain an $\varepsilon$-\textit{q}BIC (Fano-shaped optical response). (bottom) Tailored resonances that can be induced in the unpatterned film by changing different parameters of the pump beams. 
	}
  	\label{fig1}
\end{figure}

Conversely to dynamic gratings, we probe the induced permittivity modulation at wavelengths longer than the induced periodicity, for which no diffraction effect is expected (other than those at zero order) \cite{Magnusson1992, EichlerBook}. Laser beam interference has been used for imaginary refractive index-symmetry breaking and ultrafast laser output control in perovskite metasurfaces, albeit with a resonance wavelength predefined by the fabricated array of holes \cite{Huang2020}. Here, we periodically modulate the real part of the permittivity of an unstructured amorphous silicon (a-Si) film in space via a preferential carrier excitation in the bright fringes of the generated interference pattern (see Fig.~\ref{figs1} of the SI) \cite{Sjodin1998}. This modulation breaks the original continuous translational symmetry of the system and imposes a periodic 1D unit cell that is asymmetric under in-plane inversion (Fig.~\ref{fig1} top right) due to the anisotropic carrier distribution. The set of these periodic cells form an optically-induced metasurface, allowing the film to sustain {$\varepsilon$-\textit{q}BICs}. Employing this continuous, in-plane unbounded medium, we demonstrate significant optical modulation at normal incidence and ultrafast enhancement of nonlinear effects at selected wavelengths.

\section{Results}

To achieve symmetry breaking, modulations in the permittivity of a thin film need to be induced locally by the pump beams. These perturbations can arise from a range of processes, including the Kerr effect, thermal expansion, molecular reorientation and photorefraction. Some of the most significant changes can however be introduced by changing the electronic energy states, i.e. generating optically-pumped free electrons and holes in a high-index semiconductor. a-Si is a suitable material for optical excitation and can be deposited in thin films with standard equipment. To prove the feasibility of the metasurface, a 45 nm thick a-Si film was deposited on a borosilicate glass substrate (Fig.~\ref{fig1}, see Methods for fabrication details). The optically-induced permittivity modulation on it is then produced by interfering two ultrafast ($\approx$ 180 fs) laser pulses at an angle $\theta$, which produces bright and dark interference fringes with a pitch along the y-axis related to the pump wavelength $\lambda_{pump}$ via \cite{EichlerBook}:
\begin{equation}
	p_y = \frac{\lambda_{pump}}{2\sin\big(\theta/2\big)}
	\label{eq1}
\end{equation}

To reach regimes with significant changes in the film’s permittivity, these pulses with energies up to 130 nJ are focused to a $1/e^2$ waist radius of 25 $\mu$m.

\subsection{Measuring all-optical $\varepsilon$-\textit{q}BICs}

\begin{figure} 
	\centering
	\includegraphics[scale=1]{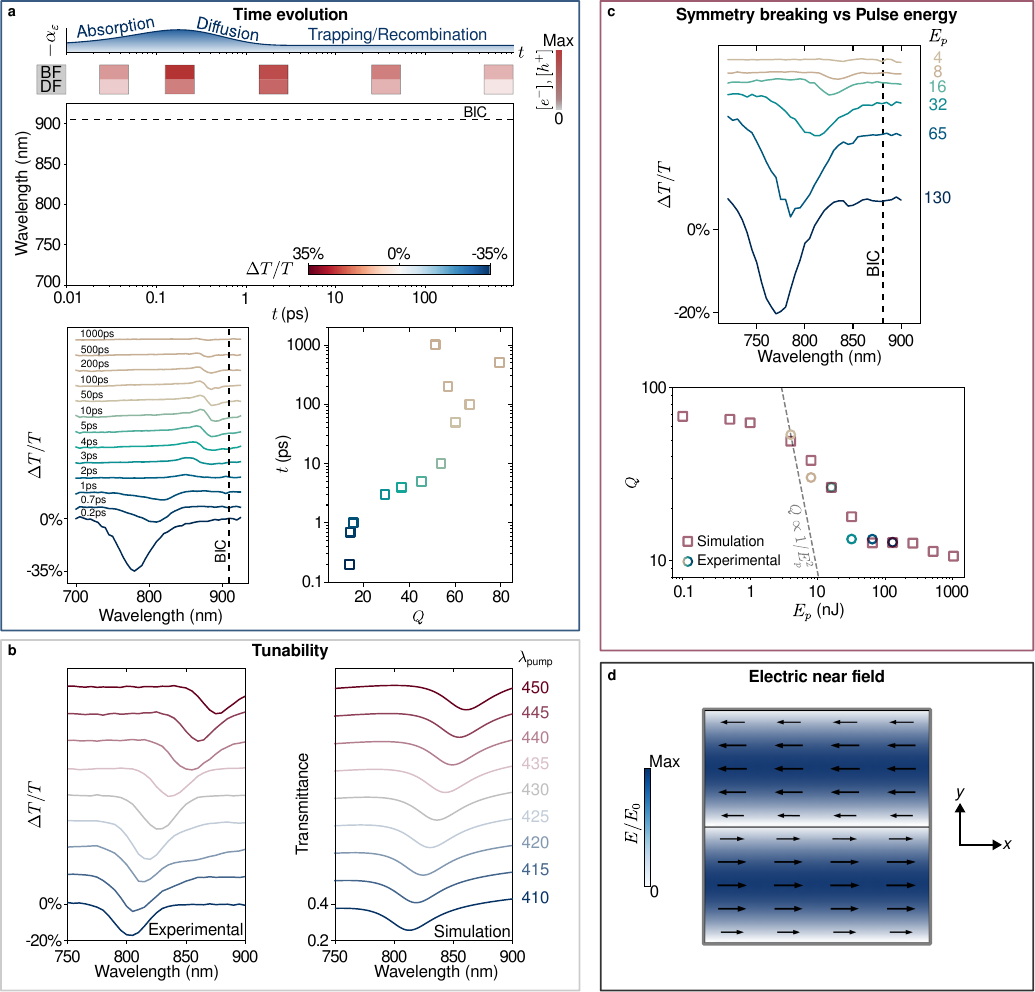}
	\caption{\textbf{Experimental all-optical $\varepsilon$-\textit{q}BICs in a featureless thin film. 
a} Temporal evolution of the $\varepsilon$-\textit{q}BIC from a 45 nm thick a-Si film measured at 130 nJ of pulse energy per arm ($\lambda_{pump} =$ 420 nm, the response of the film without a metasurface has been subtracted from the signal for clarity). Schemes on top of the differential transmittance map depict $-\alpha_{\varepsilon}$ and the free-carrier concentration ($[e^-],[h^+]$) of bright (BF) and dark fringes (DF) at different times. The resonance red-shifts towards the symmetry-protected BIC state (horizontal - on the top panel, and vertical - on the bottom left panel, dashed lines, arbitrarily positioned for guiding purposes) for longer times. Bottom right panel shows $Q$ vs $t$. $Q$-factors increase as the permittivity symmetry is restored ($\alpha_{\varepsilon} \rightarrow 0$) and losses are reduced due to carrier recombination.
\textbf{b} Tunability of $\varepsilon$-\textit{q}BICs as function of $\lambda_{pump}$. (left) Spectra measured at $\theta = 58^{\circ}$ and 40 nJ laser pulse energy per arm. (right) Corresponding spectra obtained from simulations by varying $p_y$ of the metasurface unit cell (corresponding $\lambda_{pump}$ shown on the right). Parameters $\varepsilon_{r, E_p \rightarrow \infty} =$ 0.6, \mbox{$\varepsilon^{M}_{i} =$ 0.25}, $\varepsilon_{i, E_p \rightarrow \infty} =$ 1.15, $k_b =$ 0.04/nJ, $k_d =$ 0.001/nJ and $E_p =$ 40 nJ as defined in Eq.~\ref{eq2} were used in all simulations. See Fig.~\ref{figs_modelpermittivity} of the SI for the permittivity model as a function of $E_p$.
\textbf{c} Symmetry breaking versus pump pulse energy per arm \mbox{$E_p$ (in nJ,} shown on the right, $\lambda_{pump} =$ 420 nm, see Fig.~\ref{figs3} for full differential transmittance maps). (bottom) Experimental (circles) and FEM-calculated (red squares) $Q$s vs $E_p$. An inverse-square dependence ($Q \propto 1/E_p^2$) is shown as a dashed grey line. FEM calculations employed identical parameters used in simulations of the spectra shown in \textbf{b}. 
\textbf{d} Electric near-field polarization (arrows) and electric near-field enhancement ($E/E_0$) of the \textit{quasi}-symmetry-protected resonance. A 400 nm-wide section of the metasurface along the $x$ axis is shown.}
  	\label{fig2}
\end{figure}

Since the resonances appear isolated in time after the pump pulses arrive on the substrate, the induced transmittance changes are measured with synchronized white light pulses at variable time delays. By subtracting the intrinsically flat response of the a-Si film, it is possible to spectrally resolve the time-dynamics of only the induced resonance. As seen in Fig.~\ref{fig2}a, when the white light pulse is spatially and temporally superimposed with the grating, a significant ($>$30\%) reduction of differential transmittance ($\Delta T/T$) is observed in an isolated wavelength range, which correlates to the simulated response of an induced $\varepsilon$-\textit{q}BIC under strong permittivity asymmetry (see below). This strong wavelength dependence is not expected from the theory of thin dynamic gratings \cite{EichlerBook}.

Changing the time delay of the probe pulses reveals a rich dynamic in the temporal evolution of the $\varepsilon$-\textit{q}BIC, related to the carrier dynamics in the film. The induced resonance, at first, builds up during the first 200 fs (blue-shifting to $\approx$ 780 nm) as the pump pulses are absorbed mainly at the bright fringes, progressively breaking the permittivity symmetry of the film. This resonance shows a transmittance modulation of $>30\%$ for times 0.1 ps $< t <$ 0.3 ps (see Fig.~\ref{figsultrafast} of the SI), until it slightly red-shifts and mostly disappears within approximately 1 ps. Sub-ps dynamics has been previously reported on high carrier densities in bulk crystalline Si generated via pulsed interference patterns at small angles (implying an optical grating with very large pitch, $\approx$ 13 $\mu$m) \cite{Sjodin1998}. However, relaxation times of the optical response, due to hot carriers relaxation to the band edges, had an opposite trend with pulse energy than we have observed (see below), and without any signs of strong wavelength dependence or attribution to resonant phenomena.

The decay shown in Fig.~\ref{fig2}a is attributed to the migration of excited carriers from the bright to the dark fringes, as this $\approx$ 1 ps dynamic is too fast for a full carrier recombination in the thin film (see, for example, measurements of the carrier dynamics without the metasurface in Fig.~\ref{figs3_PS} of the SI) \cite{Sokolowski2000}. This carrier migration, in turn, reduces the permittivity asymmetry of the unit cell, quantified by the asymmetry parameter $\alpha_\varepsilon \stackrel{\text{def}}{=} \Delta\varepsilon/\varepsilon$. As the red-shift continues towards the symmetry-protected BIC state (horizontal dashed line in Fig.~\ref{fig2}a top - defined by $p_y$, the substrate, the film height and unperturbed permittivity), also due to an increased refractive index attributed to carrier recombination, a second spectral feature appears. At $t \approx$ 4 ps, a long-lived resonance emerges that lasts for \mbox{$>$ 1000 ps}, attributed to trapped carriers at the bright fringes that sustain a small permittivity asymmetry (implying $\alpha_\varepsilon \neq 0$) in the metasurface.  The spectra (cross-sections of top panel) and the corresponding $Q$-factors at different times are shown in the bottom left and bottom right panels in Fig.~\ref{fig2}a, respectively.

A few aspects play a role in these dynamics. As mentioned, the excited carriers diffuse from the bright to the dark fringes, reducing the permittivity asymmetry of the system.  Smaller values of $\alpha_\varepsilon$ shift the resonance closer to the BIC state and are correlated with higher $Q$-factors \cite{Berte2023}. Besides, the excited free electrons and holes recombine and the film becomes less lossy, being able to sustain higher-$Q$ $\varepsilon$-\textit{q}BICs. This explains the appearance and red-shift of the second spectral feature observed in the differential transmittance measurements, and the respective increase in $Q$ for longer times, as shown in Fig.~\ref{fig2}a (bottom right).

To prove the spectral tunability of the metasurface, measurements were performed at different pump wavelengths, leading to different periodicities $p_y$ of the permittivity modulation (see Eq.~\ref{eq1}). The spectral response at the highest modulation for each $\lambda_{pump}$ is shown in Fig.~\ref{fig2}b, along with their respective numerically-calculated spectra. By red-shifting the pump wavelength from 410 nm to 450 nm with a fixed interference angle $\theta = 58^{\circ}$, we observe a redshift of the resonance from 805 nm to 875 nm (see Fig.~\ref{figs4} of the SI for the full differential transmittance maps). By only varying the periodicity in simulations we are able to reproduce numerically the experimental spectra in Fig.~\ref{fig2}b. The optical responses can then be calculated, with good agreement, using solely the unit cell size of the metasurface and a modified complex permittivity of a-Si as a function of the employed pump pulse energy (see below and the Numerical calculations in Methods for details). We are here neglecting the different carrier excitation efficiencies between the pump wavelengths in the formation of the metasurface.

In addition to this spectral tunability, we expect that different asymmetries lead to different $Q$-factors of the resonance. In this system, the asymmetry can be continuously adjusted by changing the incident pump pulse energy, which however also introduces different losses and changes the overall refractive index of the medium. To evaluate the response of the film and, thus, the behavior of resonances, we have therefore measured the transmission spectra at different pump pulse energies ($E_p$) between 4 nJ to 130 nJ. As seen in Fig.~\ref{fig2}c (top), higher $E_p$ values blue-shift the resonance from the assigned BIC state (vertical dashed line) and also increase their transmittance modulation (film response without the metasurface has not been subtracted). Additionally, slower decay times are observed for lower pump energies, due to the interplay between a slower carrier diffusion and lower losses due to an overall smaller excited carrier density (see the complete differential transmittance maps of the $\varepsilon$-\textit{q}BICs as a function of the pump pulse energy in Fig.~\ref{figs3} of the SI). A similar trend is observed for a pump wavelength of 410 nm (Figs.~\ref{figs_410nmpdsspectra} and \ref{figs_410nmpowerdependence} of the SI). Therefore, by merely changing the pump pulse energy not only the wavelength and amplitude of resonances can be modified, but also their $Q$-factor and decay times.

A hallmark of symmetry-protected \textit{q}BICs is the inverse-square dependence of the $Q$-factor with the asymmetry parameter $\alpha$, as derived from a perturbative treatment of the BIC state \cite{Koshelev2018, Berte2023}. This behavior occurs for a system in which the losses are dominated by radiative processes (i.e. photon emission through coupling to the continuum of states induced by geometrical/permittivity asymmetries). This implies that either non-radiative losses (absorption) are completely absent, which is an ideal case, or can be neglected when compared to radiative losses. If the $Q$-factor is described by a reciprocal sum of radiative ($r$) and non-radiative ($nr$) quality-factors ($1/Q = 1/Q_r + 1/Q_{nr}$), $Q_{nr} \rightarrow \infty$, and thus $Q = Q_r$. As the pump pulse energy is our symmetry-breaking tuning parameter, it is worthy investigating how $Q$ behaves as a function of $E_p$, and if $E_p$ can play an analogous role of geometrical \cite{Koshelev2018} or in-plane momentum perturbations \cite{Lee2012} in symmetry-protected \textit{q}BICs (i.e. $Q \propto 1/E_p^2$).

In the bottom panel of Fig.~\ref{fig2}c we show the calculated (red squares) and the experimental (circles) values of $Q$ as a function of $E_p$. The calculations for the ideal lossless metasurface versus $-\alpha_\varepsilon$ is shown in Fig.~\ref{figs_inversesquare} of the SI. As expected, when dealing with a system devoid of intrinsic losses $Q$ of the metasurface follows an inverse-square dependence with $\alpha_\varepsilon$ ($Q \propto \alpha_\varepsilon^2$), in agreement with theoretical predictions \cite{Yu2022, Berte2023, Li2023}. This corresponds to a situation where the permittivity difference between the bright and dark fringes in the 1D unit cell is gradually increased, and the losses in the grating only occur due to radiative processes as the metasurface resonance progressively couples to propagating states in the continuum. When the experimental $Q$ (see Fig.~\ref{figs_420nmpowerdependencefit} for fitted data) is plotted against $E_p$, however, an obvious deviation from a $1/E_p^2$ behavior (dashed grey line) is observed, which implies that losses are present and the asymmetry parameter $\alpha_\varepsilon$ is not a mere linear function of $E_p$ in the measured range. In spite of these deviations, we may reproduce the observed behavior by resorting to a simple model of the permittivity of bright and dark fringes as a function of $E_p$. 

\newpage
We may model the dispersive permittivity of the thin Si film ($\varepsilon^{Si}_{b(d)} (\omega, E_p)$) in the bright (dark) fringes as a function of the pump pulse energy $E_p$ as:
\begin{equation}
	\begin{split}
	\varepsilon^{Si}_{b(d)} (\omega, E_p) = 
		 \varepsilon^{Si}_{r}(\omega) \times &\bigg[\varepsilon_{r, E_p \rightarrow \infty} + (1 - \varepsilon_{r, E_p \rightarrow \infty})e^{-k_{b(d)}E_p}\bigg] \quad + \\
		& \quad i\bigg[\varepsilon^{Si}_{i}(\omega) + \varepsilon^{M}_{i} + \varepsilon_{i, E_p \rightarrow \infty}(1 - e^{-k_{b(d)}E_p})\bigg]
	\end{split}
	\label{eq2}
\end{equation}
where $\varepsilon^{Si}_{r}(\omega)$ and $\varepsilon^{Si}_{i}(\omega)$ are the real and the imaginary permittivities, respectively, of the unperturbed a-Si film as measured via ellipsometry (Fig.~\ref{figs_ellipsometry} of the SI), $\varepsilon^{M}_{i}$ a parameter that accounts for losses due to the finite size of the metasurface, $\varepsilon_{r, E_p \rightarrow \infty}$ the real (fractional) and $\varepsilon_{i, E_p \rightarrow \infty}$ the (additive) imaginary saturation values of the permittivity for $E_p \rightarrow \infty$, and $k_{b(d)}$ a constant that defines the exponential changes in the bright (dark) fringes. Three important aspects are represented in this model: (i) At very small pulse energies, $E_p \rightarrow 0$, losses are limited by the imaginary permittivity of the film and the metasurface size ($\mathfrak{Im}({\varepsilon^{Si}_{b(d)} (\omega, E_p \rightarrow 0)}) \approx  \varepsilon^{Si}_{i}(\omega) + \varepsilon^{M}_{i}$); (ii) at low pulse energies, a linear change in the permittivity is obtained ($e^{-k_{b(d)}E_p} \approx 1 - k_{b(d)}E_p$); 
(iii) at high pulse energies, $E_p \rightarrow \infty$, a saturation regime is achieved ($\varepsilon^{Si}_{b(d)} (\omega, E_p \rightarrow \infty) = \varepsilon^{Si}_{r}(\omega) \times \varepsilon_{r, E_p \rightarrow \infty} + i(\varepsilon^{Si}_{i}(\omega) + \varepsilon^{M}_{i}  + \varepsilon_{i, E_p \rightarrow \infty})$, Fig.~\ref{figs_modelpermittivity} of the SI).

This simple model is able to reproduce the trend in the measured $Q$s with a fairly good agreement. Counterintuitively, the model predicts a reduction in the asymmetry parameter ($\alpha_{\varepsilon}$) for higher pump pulse energies, as the bright fringe optical response to the pump beams saturates while the dark fringe permittivity is further reduced with increasing pump absorption (Fig.~\ref{figs_modelpermittivity} of the SI). In addition, the model predicts a saturation of $Q$ at low pulse energies ($Q \approx 70$, where losses are dominated by the finiteness of the metasurface, $\varepsilon^{M}_{i}$). Although changing $E_p$ allows us to navigate through the parameter space of asymmetries, assessing the system's response to minute values of $\alpha_{\varepsilon}$ and probing an eventual $Q \propto 1/E_p^2$ dependence requires the generation of larger metasurfaces than we are currently able with our optical setup. Besides, as the pump pulses have a Gaussian-shaped intensity distribution from its center, different carrier densities are excited at different positions in the optical 1D grating, leading to different permittivity asymmetries as we move from the center of the grating towards its edge. Naturally, by representing the metasurface using the unit cell only we are neglecting this spatial variation of $\alpha_{\varepsilon}$, nonetheless being able to reproduce the observed experimental behavior (see Fig.~\ref{figs_PvsQ_Fig2c} of the SI for the spectral response as a function of $E_p$). Further characterization of the all-optical $\varepsilon$-\textit{q}BICs (pump and probe polarization dependence, etc) can be found in Figs.~\ref{figs2}-\ref{figs_probe_polarization_CS} of the SI.

\subsection{Spectrally-selective third harmonic generation enhancement via $\varepsilon$-\textit{q}BICs}
$\varepsilon$-\textit{q}BICs excited via optically-induced metasurfaces are expected to carry the same functionalities of resonator-based metasurfaces. One of the most utilized properties of metasurfaces featuring resonances is their optical near-field enhancement, which can be used to boost phenomena that require strong light-matter interaction, like nonlinear effects \cite{Tonkaev2023, Zograf2022}. Symmetry (centrosymmetry) breaking via DC electric fields has been long used to induce second-harmonic generation in centrosymmetric materials \cite{Terhune1962}, although it is not a spectrally-selective method. As a proof-of-principle, we demonstrate how the induced in-plane symmetry breaking can be used to enhance the third harmonic generation (THG) at specific wavelengths in a bare silicon film within a narrow timeframe.

\begin{figure} [t] 
	\centering
	\includegraphics[scale=1]{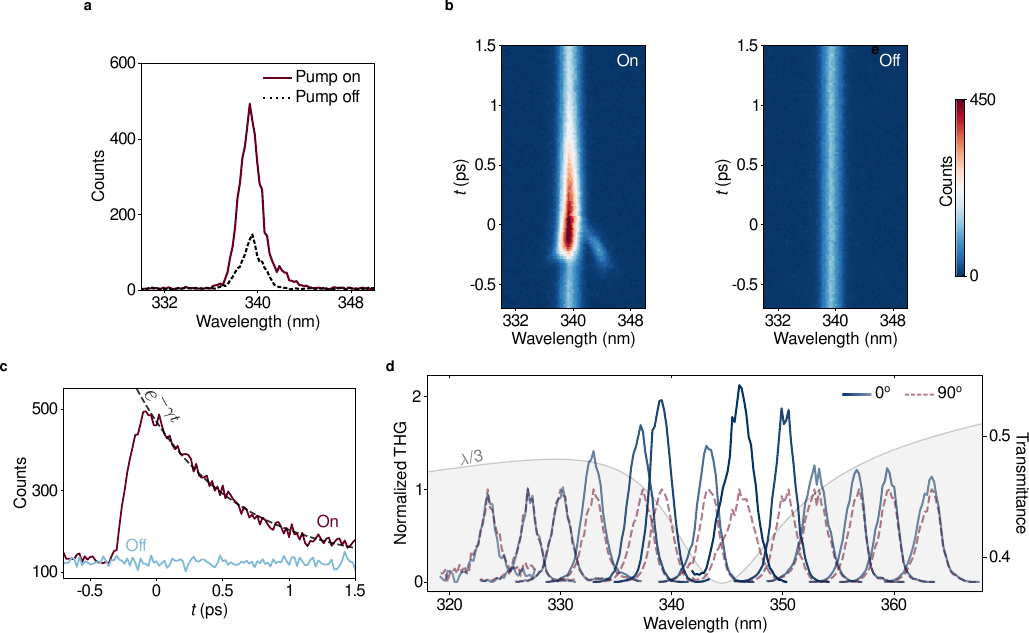}
	\caption{\textbf{Spectrally-selective third-harmonic generation enhancement using {$\varepsilon$-\textit{q}BICs}.} 
\textbf{a} THG signal (\mbox{$E_p =$ 75 nJ}) in the presence (pump on) and absence (pump off) of the optical grating ($\lambda_{pump} =$ 515 nm) providing the permittivity symmetry breaking. A 234\% enhancement of THG (at 340 nm, $\lambda_{fundamental} =$ 1020 nm) is obtained when a metasurface is generated in the thin film (solid red line) compared to when both pump beams are blocked (dashed black). 
\textbf{b} THG signal vs delay line position, whose cross-sections in time are shown in panel \textbf{a}. 
\textbf{c} Corresponding THG signals as function of time at \mbox{339 nm}. The enhancement provided by the $\varepsilon$-\textit{q}BIC decays exponentially ($1/e \approx$ 0.72 ps, fit shown as a grey dashed line). See Fig.~\ref{figs_thg_Fig3a} for full data.
\mbox{\textbf{d} Normalized THG signal for} $0^{\circ}$ (blue curves) and $90^{\circ}$ (dashed light red) relative pump polarizations as a function of the fundamental beam wavelength (from $\lambda_{fundamental} =$ 970 nm to $\lambda_{fundamental} =$ 1090 nm). A preferential THG enhancement is observed when the fundamental is overlapped with the $\varepsilon$-\textit{q}BIC (transmittance shown as a gray shadowed area, right-hand side axis). Parameters $\varepsilon_{r, E_p \rightarrow \infty} =$ 0.75, $\varepsilon^{M}_{i} =$ 0.25, $\varepsilon_{i, E_p \rightarrow \infty} =$ 1.05, $k_b =$ 0.03/nJ, $k_d =$ 5$\times 10^{-4}$/nJ, \mbox{$E_p =$ 50 nJ} and $\theta =$ 51$^{\circ}$ (implying $p_y =$ 598 nm) were used in the transmittance calculation.
A $>$ 100\% enhancement is obtained for \mbox{$\lambda_{fundamental} =$ 1040 nm.} $\varepsilon$-\textit{q}BIC spectrum was shifted to a third of its value ($\lambda/3$) to overlap with THG signals.}
  	\label{fig3}
\end{figure}

For that, we generate a grating at 515 nm pump, which we now probe with pulses of tunable narrow bandwidth (fundamental beam) instead of the supercontinuum white light. The longer pump wavelength increases the $p_y$ of the metasurface, red-shifting the $\varepsilon$-\textit{q}BIC further into the infrared range, and therefore its THG closer to the visible. In Fig.~\ref{fig3}a the THG response of the bare film around 340 nm \mbox{($\lambda_{fundamental} =$ 1020 nm)} is compared when both pump beams are either turned on (solid red) or turned off (dashed black). A 234\% enhancement of the THG signal is obtained when the pump beams are illuminating the sample and breaking the in-plane inversion permittivity symmetry of the thin film. The corresponding THG signals as a function of time are shown in Fig.~\ref{fig3}b and Fig.~\ref{fig3}c. The spot at \mbox{$\approx$ 343 nm} in Fig.~\ref{fig3}b is due to a four-wave mixing process, disappearing in the absence of the pump beams (Fig.~\ref{fig3}c, see SI for further characterization). The decay of the THG enhancement clearly follows an exponential trend (fitted grey dashed line) instead of a Gaussian temporal profile expected from an enhancement due to a mixing of ultrafast beams. This $\approx$ 1 ps dynamic is temporally similar to the excitation of $\varepsilon$-\textit{q}BICs as shown in Fig.~\ref{fig2}a.

To evaluate the spectral response of the nonlinear effect, sweeps at different probe wavelengths were taken and referenced with sweeps where the pump pulses were cross-polarized, which switched off the grating contrast. Given the different collection efficiencies at different wavelengths of our optical setup, we have normalized the THG output (see Figs.~\ref{figs_thg_lambda}-S28 in the SI for the raw data). By comparing the THG signals with ($0^{\circ}$) and without ($90^{\circ}$) the metasurface, we observed a preferential enhancement for probe wavelengths that are spectrally overlapped with the $\varepsilon$-\textit{q}BIC (grey shadowed area in Fig.~\ref{fig3}d, whose spectra was shifted to $1/3$ of its value to overlap with the THG output). Here, a \mbox{$>$ 100\%} enhancement was obtained for $\lambda_{fundamental} =$~1040~nm ($\lambda_{THG} \approx $ 347 nm) in the presence of the metasurface, when compared to the $90^{\circ}$-rotated polarization of one of the pump beams, despite all incident radiation of pump and probe having the same pulse energy in both situations. The lower enhancement is attributed to the smaller pump energy used for this set of measurements ($E_p$ = 50 nJ vs $E_p$ = 75 nJ in Fig.~\ref{fig3}a).

\section{Discussion}

We have shown that a simple unpatterned dielectric amorphous film can yield tailored resonances when its permittivity is periodically modulated in space, breaking its original translational and in-plane inversion symmetries. Optical resonances do not require then to be pre-defined and constrained by etching and deposition fabrication procedures that provide the boundary conditions for typical spatially-finite resonators. $\varepsilon$-\textit{q}BICs demonstrated here in the near infrared can easily be extended further into the visible and towards the mid- and far-infrared, where coupling to excitonic resonances \cite{Weber2023} and molecular vibrational/rotational levels could be performed \cite{Xie2023, Tittl2018}.

The emergence of the symmetry-protected \textit{quasi-}BICs in the unbounded medium cannot be easily, nor intuitively, derived from the optical properties of the thin film, i.e., regardless how much one knows about its atomic and electronic band structure that define its optical properties, the symmetry-protected \textit{q}BIC does not follow trivially from this knowledge \cite{Anderson1972}.

Periodic permittivity variations in space are also a core concept in distributed Bragg reflectors (DBRs) and photonic crystals (PhCs) \cite{Rayleigh1888, Yablonovitch1987, John1987}. But, conversely to DBRs and PhCs, where confinement of light deteriorates with a smaller contrast of refractive indices, here higher localization and, thus, higher $Q$-factors are achieved with smaller spatial permittivity differences. This facilitates the dynamic modulation of devices, as it reduces - or, in principle, completely circumvents - the need for large alterations in optical properties of materials with a given input to significantly change their optical response. The minute values of electro-optical coefficients of most materials here play to our advantage, and a number of substances which were previously dismissed as ``optically inert" can now be incorporated into the set of electro-optical materials, including silicon. Besides free-carriers, any permittivity-change mechanism can be employed for symmetry breaking and the excitation of $\varepsilon$-\textit{q}BICs in a thin film, be it via linear (Pockels) \cite{Abel2019} or nonlinear (Kerr) electric-field effects \cite{Boydbook}, ion implantation \cite{Basu2010}, temperature differences \cite{Plotnik2011}, etc.

The broad optical responses in condensed matter imply that any given input aimed to modulate the permittivity of materials would also affect it in a broadband fashion \cite{Fox2002, Soref1987}. For this reason, optical signals that are encoded in different frequencies within an optical fiber (wavelength-division multiplexing) need to be physically separated in demultiplexing operations. By leveraging the strong and spectrally-localized resonances that arise from small perturbations in permittivity symmetries in an otherwise optically-flat platform, we may modulate/filter the propagation of a small set of wavelengths independently, leaving the propagation of others frequencies nearly undisturbed. Selective optical modulation using $\varepsilon$-\textit{q}BICs across all optical telecommunication bands (from the high-energy edge of the \mbox{O band} at 1260 nm to the low-energy end of the L band at 1625 nm) can, in principle, be easily performed in the same thin silicon film. 

As we have demonstrated, several degrees of freedom can be used to tailor the desired resonances, such as pump pulse energy and wavelength, relative delay, incidence angle and polarization of the grating-forming beams. Another advantage of a homogeneous medium relative to pre-fabricated metasurfaces of resonators is the absence of losses due to corners, edges and fabrication imperfections. This makes defect-free crystalline films a sort of ultimate platform for on-demand high-$Q$ modes. We have shown the excitation of $\varepsilon$-\textit{q}BICs through a rather lossy symmetry-breaking mechanism, although it allowed us to exploit large values of the asymmetry parameter $\alpha_\varepsilon$ and a way to assess ultrafast carrier dynamics and spatial distribution in solid state systems \cite{EichlerBook}. Besides, selectively enhancing nonlinearities in a platform whose frequency-conversion efficiency is not constrained by phase-matching requirements, but instead governed by the induced resonances, provides great flexibility when exploiting nonlinear phenomena. Crucial questions concern what practical limits this platform can achieve, and what other asymmetries may be exploited \cite{Liu2020, Yu2022, Huang2020}. The features of active metasurfaces may be expanded to include permittivity symmetry breaking in continuous media.

\section{Conclusion}

In conclusion, we have demonstrated that an unbounded, featureless thin film with an intrinsically flat optical response can sustain resonances at desired wavelengths, amplitudes and $Q$-factors when a particular permittivity asymmetry is induced in it via optical means, i.e., by interfering two ultrafast laser beams at an angle. We have argued that the nontrivial emergence of strong, frequency-dependent, optical responses follows directly from the symmetry breaking of the original continuous translational and in-plane inversion permittivity symmetries of the film.

Experiments were performed in one of the simplest CMOS-compatible platforms, namely, an amorphous Si (a-Si) film over borosilicate glass. Transmittance modulations as high as 35\% and $Q \approx 80$ for long-lasting resonances (observed due to carrier trapping) were achieved via carrier excitation in the semiconductor, an inherently lossy symmetry-breaking mechanism. Significant improvements are expected in modulation amplitudes and in $Q$-factors of $\varepsilon$-\textit{q}BICs if a crystalline Si film, inherently less lossy, is used instead, by pumping at longer wavelengths and also by exploiting larger metasurfaces than we have used here. We have also demonstrated a sub-ps $>$ 200\% enhancement of THG in the bare Si film at wavelengths corresponding to the $\varepsilon$-\textit{q}BIC excitation. Again, the inherently lossy mechanism of symmetry breaking due to carrier excitation inhibits larger enhancements of the third harmonic than we have observed, a process whose efficiency should scale enormously for larger $Qs$ (proportional to the dwelling time of the photons in the film) and due to increasing \mbox{near-field enhancement.} 

We believe that $\varepsilon$-\textit{q}BICs in featureless continuous media have a strong connection with GMRs in unpatterned thin films \cite{Wang1990, Magnusson1992, Wang1993}. Indeed, a $Q \propto 1/\alpha^2(\theta)$ dependence and the symmetry-protected nature of GMRs have been observed in silicon cuboid metasurfaces at THz frequencies due to an incidence angle($\theta$)-induced symmetry breaking \cite{Han2019}. However, while GMRs are essentially invariant in one dimension, all-optical $\varepsilon$-\textit{q}BICs of higher dimensionality could be excited \cite{Liu2020, Yu2022, Berte2023, Li2023, Huang2020}. Besides, GMRs have required so far featureless thin films to be coupled to corrugated structures in resonant waveguide gratings \cite{Quaranta2018} or to bear inclusions in high-contrast gratings \cite{Sturmberg2015}, both already exploited in several applications. The addition of corrugations or inclusions, however, inevitably defines the periodicity of the structure and thus its resonant wavelength, limiting the spectral flexibility of the design. All-optical $\varepsilon$-\textit{q}BICs here demonstrated do not have this structural limitation.

We hope that the all-optical excitation of $\varepsilon$-\textit{q}BICs will provide new ways of dynamically manipulating electromagnetic radiation at required wavelengths in a single lithography-free platform and bring new possibilities for light technologies. These may encompass applications as varied as optical filters \cite{Magnusson1992}, $Q$-switching \cite{McClung1967}, holography, focusing and steering of optical beams \cite{Malek2020} and also thermal emission control \cite{OvervigPRX2021} through the interplay of local and nonlocal responses, chiral \cite{Overvig2021} and orbital angular momentum \cite{Wang2020} manipulation, generation of second- and higher-order harmonics \cite{Tonkaev2023, Zograf2022}, optical modulators \cite{Reed2010}, tunable lasing \cite{Kodigala2017, Noda2001, Hirose2014, Huang2020}, sensing \cite{Tittl2018}, coupling to vibrational \cite{Xie2023} and excitonic resonances \cite{Weber2023} in controllable weak to ultrastrong light-matter interaction regimes \cite{Kockum2019} and polaritonic condensation \cite{Berghuis2023}.

\newpage

\section{Methods}
\label{methods}
\subsection{Numerical calculations}

Thin film spectra were calculated through finite-difference time domain solutions of Maxwell’s equations using the commercially available software Lumerical (Ansys). A normal incidence, linearly-polarized (along the $x$-axis) plane wave propagating in the $-z$ direction illuminated the thin film. Periodic boundary conditions were used to represent the in-plane ($xy$) unbounded media. Power monitors were used for transmittance calculations, and perfectly matched layers (PMLs) domains at the top and bottom for the absorption of propagating waves. A dispersionless refractive index of $n=1.45$ was employed for the borosilicate substrate, while the dispersive complex permittivity of the amorphous silicon (a-Si), as measured via ellipsometry was used for the thin film (see Supplementary Information for data, Fig.~\ref{figs_ellipsometry}). The dispersive complex permittivity of a-Si was then modified with the model proposed in Eq.~\ref{eq2} for bright and dark fringes (Fig.~\ref{figs_modelpermittivity} of the SI). The modified permittivity was considered homogeneous in each fringe. Eigenfrequency solutions of Maxwell’s equations were used for $Q$-factor calculations and were performed using the commercially available RF module of the finite element solver COMSOL Multiphysics with identical boundary conditions as described above. In both types of simulations we have represented the optically-induced metasurface using only the unit cell (Fig.~\ref{fig1}, top right). This approximation reduces the computational resources required for simulating and predicting the desired optical responses of the thin film when compared to a simulation of the full grating domain \mbox{(tens of $\mu m \times$tens of $\mu m$).} By simulating the unit cell alone we have then neglected spatial permittivity variations from the center towards the outer edges of the metasurface, which would be inevitably induced by the Gaussian intensity profiles of the pump beams used to generate the grating.

\subsection{Sample fabrication}

a-Si was deposited onto borosilicate glass substrates via PECVD. Samples were annelead at 700$^{\circ}$ C for 90 s to improve film quality.

\subsection{Optical measurements}

A collinear optical parametric amplifier (OPA, ORPHEUS-HP) pumped by a pulsed Yb:KGW Pharos laser system (Light Conversion Ltd) of maximum repetition rate of 200 kHz generated outputs of pulse duration $\approx$ 180 fs. For time-dependent transmittance measurements, the 515 nm invariant beam output of the OPA was focused onto a 5-mm-thick sapphire plate and used to generate the supercontinuum probe light beam (see spectrum in the SI, Fig.~\ref{figs_whitelight}), while the mechanically-chopped wavelength tunable OPA output was sent through a 50/50 beam splitter to generate the metasurface. A motorized optical delay line was used to introduce controlled time differences between the pump and supercontinuum pulses. One of the pump beams passed through a controlled half-wave plate and a neutral density wheel for adjustable attenuation. The grating-generating pump beams and the supercontinuum light were slightly focused onto the sample using lenses of focal length equal to 100mm and 75mm, respectively. Transmittance measurements were carried out with a lock-in detection system (Stanford Research Instruments) by modulating the wavelength tunable OPA output at $<2$ kHz frequency using an optical chopper. A spectrograph (PI Acton SP2300, Princeton Instruments) coupled to an avalanche photodiode (Thorlabs APD440A) was used for spectral characterization of the supercontinuum probe light transmitted by the sample. THG measurements were performed by using the invariant 515 nm output of the OPA to generate the metasurface, while the variable output (previously used as the pump beams) was used as the fundamental beam to generate the third harmonic response in the thin film.

\section{Data availability}

The data that support the findings of this study are available from the corresponding author upon reasonable request.

\section{Acknowledgments}

We acknowledge financial support from the Deutsche Forschungsgemeinschaft (DFG, German Research Foundation) under Grant Nos. EXC 2089/1 - 390776260 (Germany's Excellence Strategy and TI 1063/1 - Emmy Noether Program), the Bavarian State Ministry of Science, Research, and Arts through the program “Solar Technologies Go Hybrid (SolTech)” and the Enabling Quantum Communication and Imaging Applications (EQAP) project. Funded by the European Union (ERC, METANEXT, 101078018). Views and opinions expressed are however those of the author(s) only and do not necessarily reflect those of the European Union or the European Research Council Executive Agency. Neither the European Union nor the granting authority can be held responsible for them. S.A. Maier additionally acknowledges the Lee-Lucas Chair in Physics and the Australian Research Council. We thank local research clusters and centers such as the Center of Nanoscience (CeNS) for providing communicative networking structures.

\section{Authors contributions}

R.B. conceived the project and performed the theoretical and numerical analysis. R.B., T.P. and L. de S.M. conceived the experimental demonstrations. T.P. and L. de S.M. assembled the optical setup and performed the measurements. R.B. and T.P. performed the experimental data analysis.  R.B., L. de S.M., A.T. and S.A.M. supervised the project. R.B. wrote the manuscript. All other authors contributed to the manuscript improvement.

\section{Competing interests}

The authors declare no competing interests.

\section{Additional information}

Supplementary Information is available for this paper. Correspondence and requests for materials should be addressed to  R.B. (R.Berte@physik.uni-muenchen.de) and A.T. (A.Tittl@physik.uni-muenchen.de).

\bibliography{library}
\bibliographystyle{ieeetr}

\xdef\presupfigures{\arabic{figure}}
\renewcommand\thefigure{S\fpeval{\arabic{figure}-\presupfigures}}


\linespread{1.72} 


\title{Supplementary Information for \\ Emergent resonances in a thin film tailored by optically-induced small permittivity asymmetries}

{
\let\clearpage\relax
\maketitle
}

\newpage

\newpage

\section{Optical setup}

\begin{figure}[h] 
	\centering
	\includegraphics[scale=1.3]{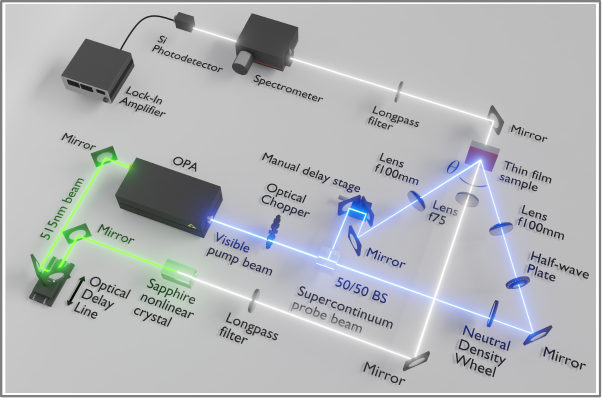}
	\caption{Scheme of the optical setup employed in the generation of the optically-induced metasurface and detection of {$\varepsilon$-\textit{q}BICs}. The spectrum of the supercontinuum probe beam is shown in Fig. S2.}
  	\label{figs_optical_setup}
\end{figure}

\vspace{10ex}

\section{Spectrum of the supercontinuum probe beam}

\begin{figure}[h] 
	\centering
	\includegraphics[scale=1]{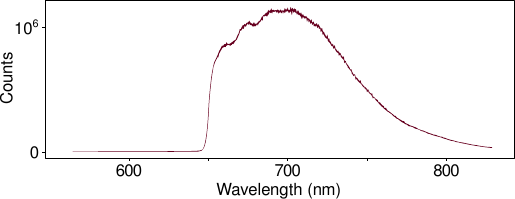}
	\caption{Spectrum of the supercontinuum probe beam generated from the 515 nm OPA output ($P_{515nm} =$ 210 mW) in a sapphire nonlinear crystal. The spectrum was filtered with a long-pass 650 nm optical filter prior to measurements. The generated supercontinuum beam is heavily chirped, which has been manually corrected in the post-processing of data for analysis when necessary.}
  	\label{figs_whitelight}
\end{figure}

\newpage

\section{Optical response versus path (time) delay between pump pulses}

\begin{figure}[h] 
	\centering
	\includegraphics[scale=1]{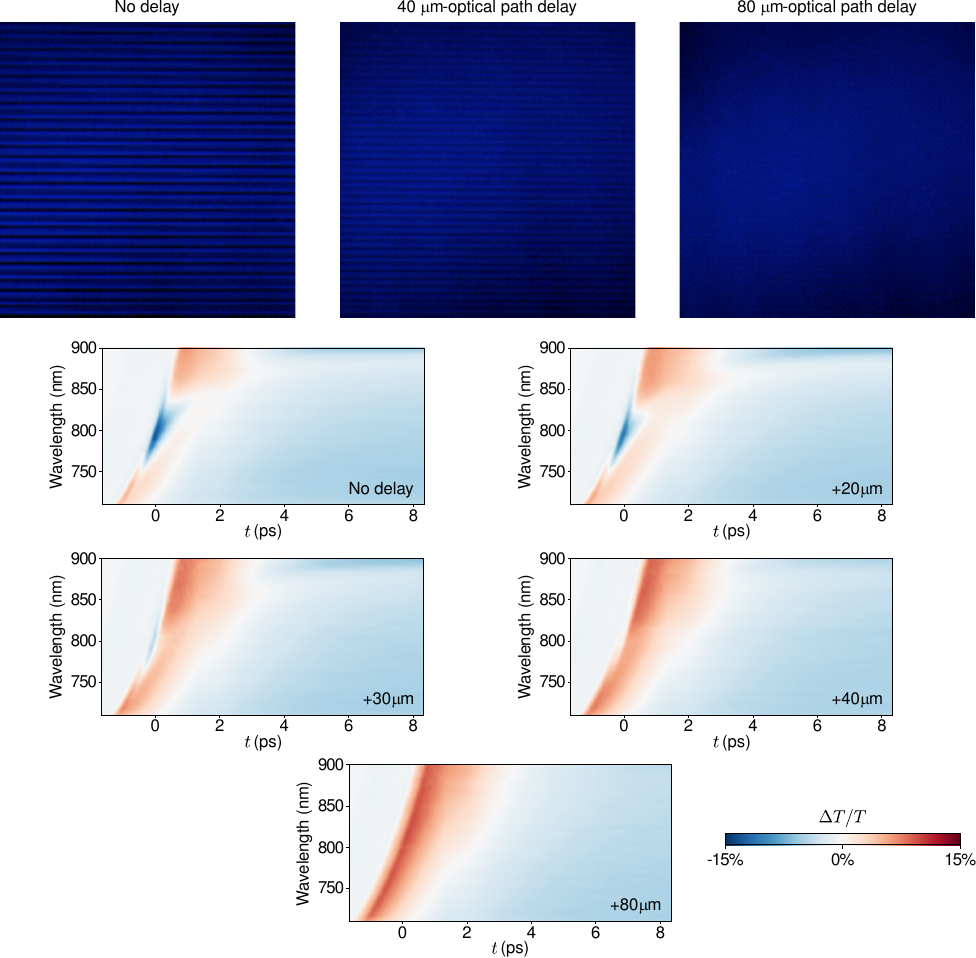}
	\caption{Optical micrographs ($\lambda=430 nm$) without (left), and with temporal delays between pulses induced by a 40 $\mu$m (or 40 $\mu$m/3$\times10^8$ m/s = 133 fs, center) and by an 80 $\mu$m (or 266 fs, right) increase in the optical path of one of the pump beams. The 1D optical grating progressively disappears as the pulses arrive in the thin film at increasingly different times. (bottom) Temporal evolution of $\varepsilon$-\textit{q}BICs as a function of the optical path delay (shown in the insets). As the 1D optical grating disappears, so does the induced resonance, and only the flat optical response of the thin film is observed.}
  	\label{figs1}
\end{figure}

\newpage

\section{Temporal cross sections of $\varepsilon$-\textit{q}BICs}

\begin{figure}[h] 
	\centering
	\includegraphics[scale=1]{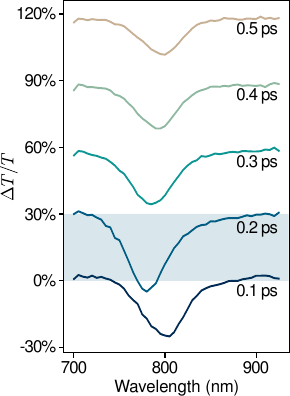}
	\caption{Temporal cross-sections of data shown in Fig. 2a of the main manuscript, for 0.1 ps $< t <$ 0.5 ps. Although the main resonance lasts for $\approx$ 1 ps, the transmittance modulation is larger than 30\% (shadowed area) only between 0.1 ps and 0.3 ps. The cross-sections at different probe delays are shifted by 30\% for better visualization.}
  	\label{figsultrafast}
\end{figure}

\vspace{0ex}

\section{Optical response of the film versus pump pulse energy in the absence of the optically-induced metasurface}

\begin{figure}[h] 
	\centering
	\includegraphics[scale=1]{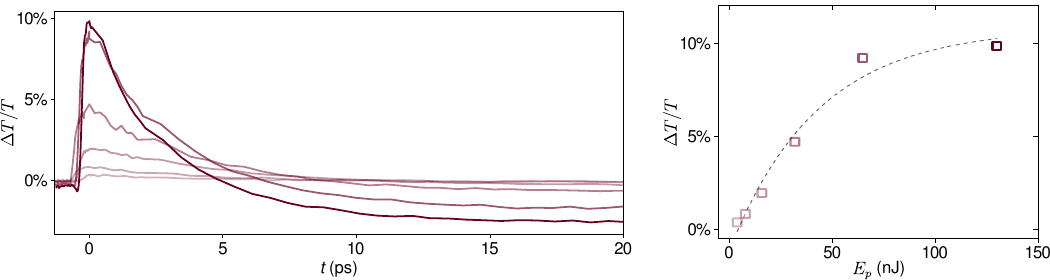}
	\caption{(left) Differential transmissivity at 800nm as a function of time for a 90{$^{\circ}$}-rotated relative polarization between the pump beams (a condition where no 1D optical grating, thus no metasurface, is formed). Increasing pump pulse energies are shown in darker red colors. $\lambda_{pump} =$ 420 nm. (right) Respective maximum differential transmissivity vs pump pulse energy per arm (at $t =$ 0 ps). Dashed line shows the exponential fit ($ \Delta T/T = a_0 + a_1e^{-kE_p}$) of the data. The saturation of the optical response and a deviation from a linear behavior are clearly observed for \mbox{pulse energies per arm larger than 65 nJ.}}
  	\label{figs3_PS}
\end{figure}

\newpage

\section{Temporal evolution of $\varepsilon$-\textit{q}BICs as a function of the pump wavelength}

\begin{figure}[h] 
	\centering
	\includegraphics[scale=1]{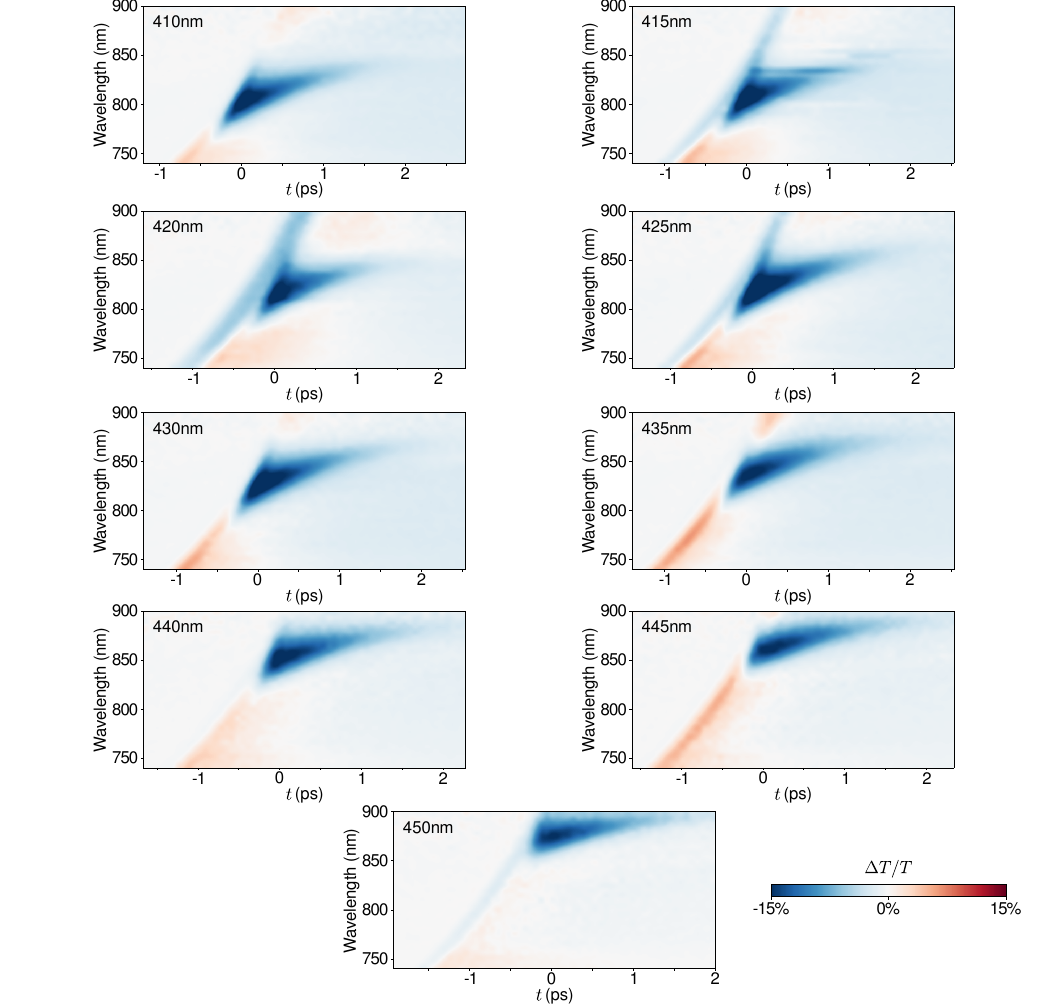}
	\caption{Temporal evolution of $\varepsilon$-\textit{q}BICs as a function of the pump wavelength (insets). Resonances red-shift for longer pump wavelenghts, corresponding to larger periodicities $p_y$ of the metasurface. All resonances have a similar temporal profile, lasting for $\approx$ 1 ps and shifting towards the symmetry-protected BIC state for longer times, as the permittivity symmetry is restored due to carrier migration/recombination. Film response without the metasurface was subtracted to highlight resonances.}
  	\label{figs4}
\end{figure}

\newpage

\section{Optical grating periodicity versus pump wavelength}

\begin{figure}[h] 
	\centering
	\includegraphics[scale=1]{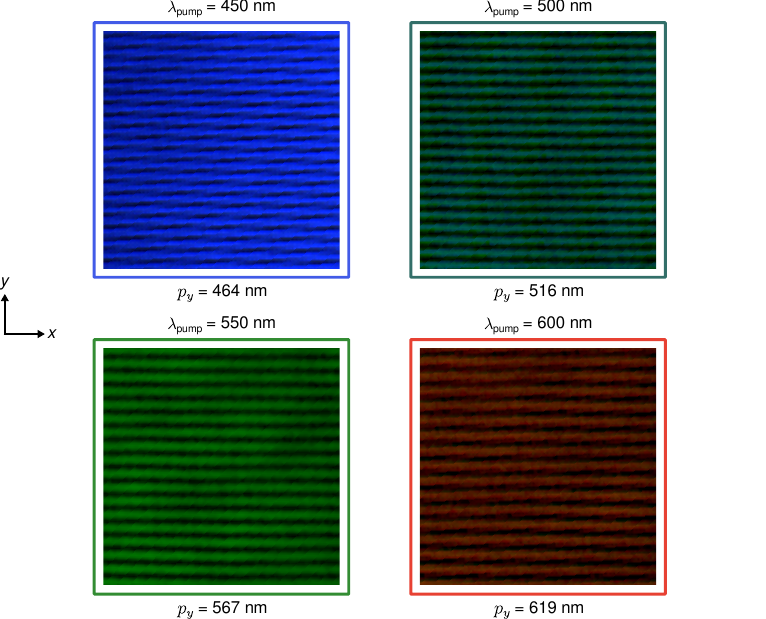}
	\caption{Micrographs of optical 1DGs: larger periodicities ($p_y$) are notably formed for longer excitation wavelengths $\lambda_{pump}$ for the same incidence angle $\theta = 58^{\circ}$.}
  	\label{figs_1DGvsLambda}
\end{figure}


\newpage

\section{Optical response versus the pump pulse energy}

\begin{figure}[h] 
	\centering
	\includegraphics[scale=1]{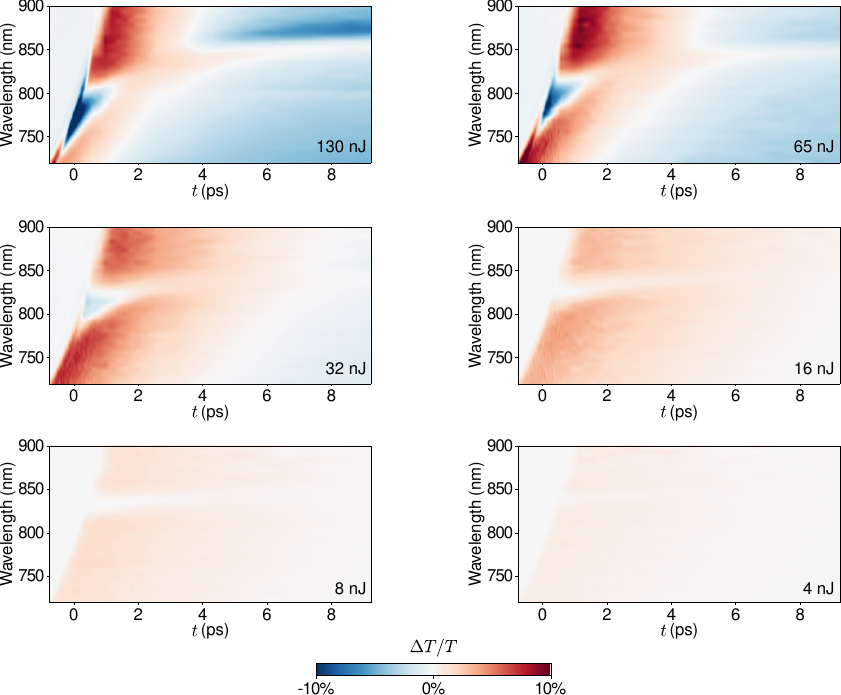}
	\caption{Temporal evolution of $\varepsilon$-\textit{q}BICs as a function of the pump pulse energy per arm (indicated in the insets), \mbox{$\lambda_{pump} =$ 420 nm}. Resonances red-shift towards the symmetry-protected BIC state as the pump power is reduced.
	}	
  	\label{figs3}
\end{figure}

\newpage

\begin{figure}[h] 
	\centering
	\includegraphics[scale=1]{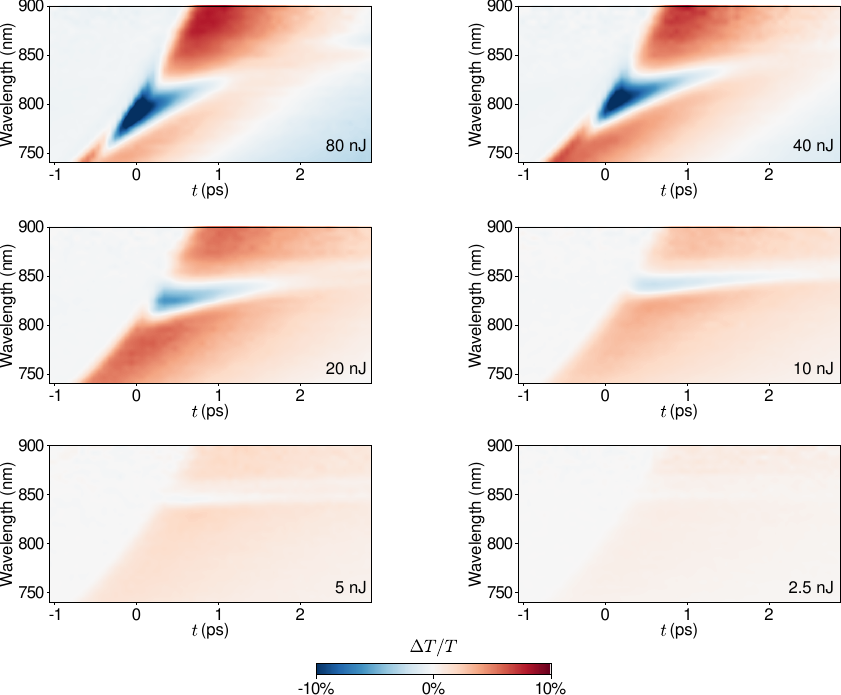}
	\caption{Temporal evolution of $\varepsilon$-\textit{q}BICs as a function of the pump pulse energy per arm (insets), $\lambda_{pump} =$ 410 nm. Similar to the $\lambda_{pump} =$ 420 nm case, resonances red-shift towards the symmetry-protected BIC state and decrease in amplitude modulation as the pump power is reduced.}
  	\label{figs_410nmpdsspectra}
\end{figure}

\newpage

\begin{figure}[h] 
	\centering
	\includegraphics[scale=1]{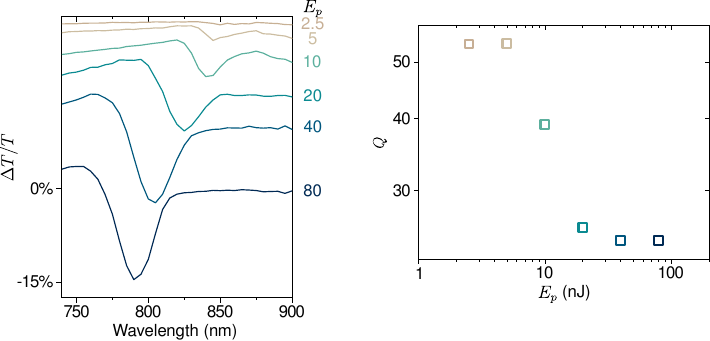}
	\caption{Spectra (left) and respective $Q$-factors (right) of $\varepsilon$-\textit{q}BICs as a function of the pump pulse energy per arm (\mbox{$E_p$ in nJ}, indicated on the right hand side of the corresponding spectrum), $\lambda_{pump} =$ 410 nm. Resonances blue-shift away from the symmetry-protected BIC state and increase in transmittance modulation for larger pump powers, being 80 nJ per pump arm the upper power limit of our setup for this wavelength.}
  	\label{figs_410nmpowerdependence}
\end{figure}

\vspace{15ex}

\section{$Q$ versus the asymmetry parameter for a lossless metasurface}

\begin{figure}[h] 
	\centering
	\includegraphics[scale=1.1]{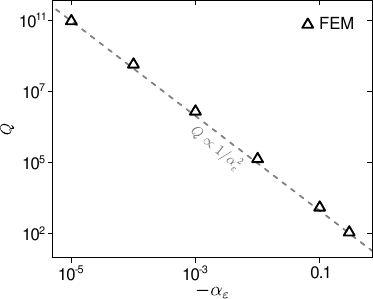}
	\caption{Finite element method (FEM)-calculated $Q$ of a lossless metasurface (triangles) vs the negative (due to a decreasing permittivity with increasing free-carrier concentration) of the asymmetry parameter $\alpha_\varepsilon$, in agreement with the theoretical prediction (dashed grey line, $Q \propto 1/\alpha_\varepsilon^2$).}
  	\label{figs_inversesquare}
\end{figure}

\newpage

\section{Fitting of differential transmittance data}

\begin{figure}[h] 
	\centering
	\includegraphics[scale=1]{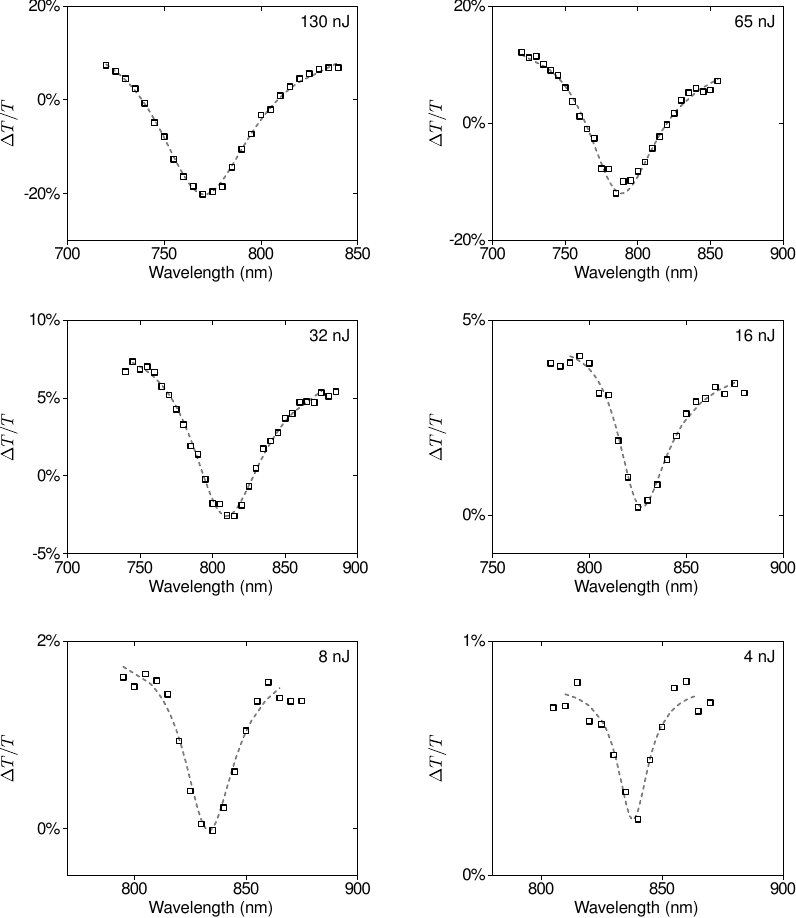}
	\caption{Fitting of differential transmittance for different pump pulse energies (insets), $\lambda_{pump} =$ 420 nm. A Breigt-Wigner-Fano (BWF) function $\frac{\Delta T}{T} = T_0 + \frac{H\big(1+\frac{\lambda-\lambda_c}{qw}\big)^2}{1+\big(\frac{\lambda-\lambda_c}{w}\big)^2}$ was used to fit (dashed lines) the data (open squares), where $H$ is the height, $q$ is the Fano asymmetry parameter, $\lambda_c$ is the central wavelength of the resonance and $w$ its width. The resulting $Q$-factor is calculated as $Q = \lambda_c/2w$. 
	}	
  	\label{figs_420nmpowerdependencefit}
\end{figure}

\newpage

\begin{figure}[h] 
	\centering
	\includegraphics[scale=1]{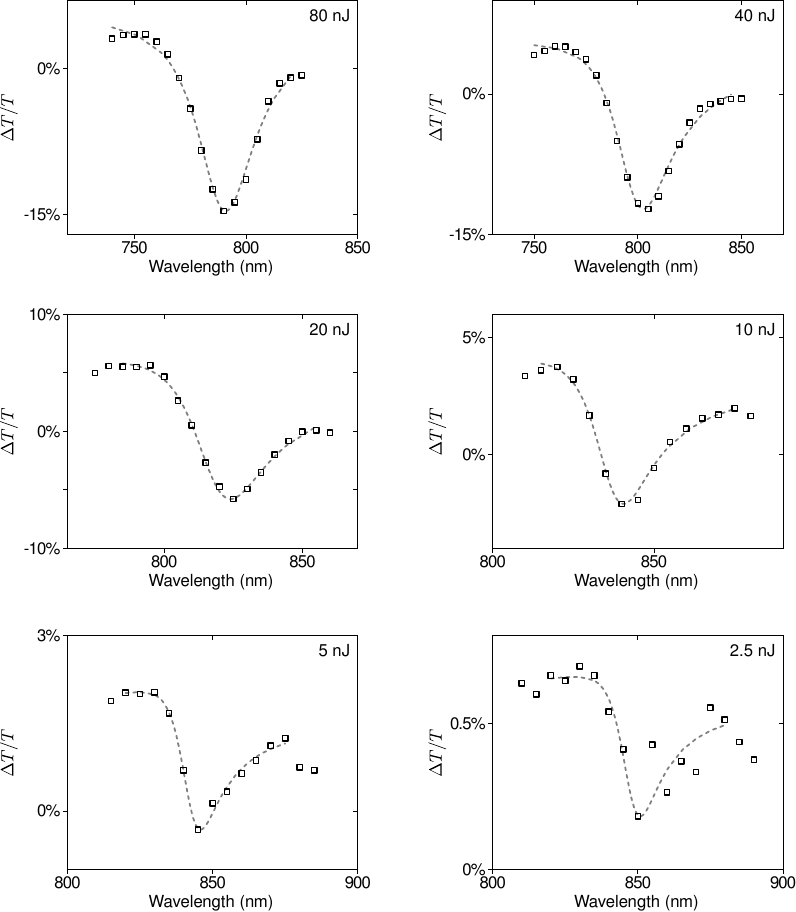}
	\caption{Fitting of differential transmittance for different pump pulse energies (insets), $\lambda_{pump} =$ 410 nm. A Breigt-Wigner-Fano (BWF) function was used to fit (dashed lines) the data (open squares). 
	}	
  	\label{figs_410nmpowerdependencefit}
\end{figure}

\newpage

\section{Permittivity data of silicon films}

\begin{figure}[h] 
	\centering
	\includegraphics[scale=1]{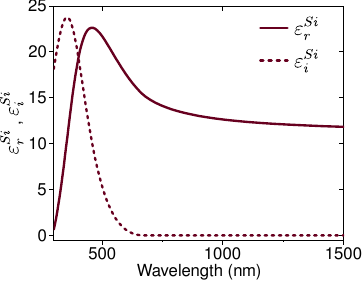}
	\caption{Real ($\varepsilon_r^{Si}$, solid line) and imaginary ($\varepsilon_i^{Si}$, dashed line) parts of the permittivity of the fabricated a-Si films as a function of the wavelength, measured via ellipsometry.}
  	\label{figs_ellipsometry}
\end{figure}

\vspace{15ex}

\section{Permittivity and permittivity asymmetry as a function of the pump pulse energy}

\begin{figure}[h] 
	\centering
	\includegraphics[scale=1]{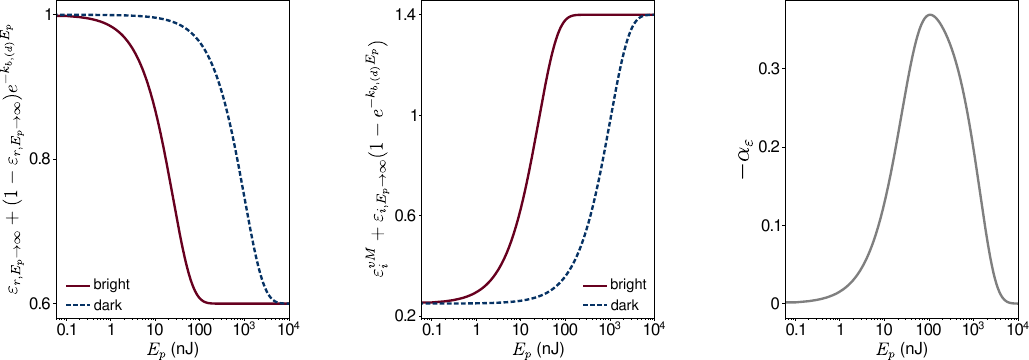}
	\caption{Real (multiplicative, left) and imaginary (additive, center) permittivity terms of bright and dark fringes of the metasurface as a function of the pump pulse energy per arm ($E_p$). (right) $-\alpha_{\varepsilon}$ versus $E_p$. Values were obtained from the model shown in \mbox{Eqn. 2} of the main manuscript. The parameters values used in model are $\varepsilon_{r, P \rightarrow \infty} = 0.6$, $\varepsilon^{vM}_{i} = 0.25$, $\varepsilon_{i, P \rightarrow \infty} = 1.15$, $k_{b} = 0.04$/nJ and $k_{d} = 0.001$/nJ. $-\alpha_{\varepsilon}$ decreases for large pulse energies due to an absorption saturation of the bright fringes.
	}
  	\label{figs_modelpermittivity}
\end{figure}

\newpage

\section{Spectrum versus pump pulse energy}

\begin{figure}[h] 
	\centering
	\includegraphics[scale=1.1]{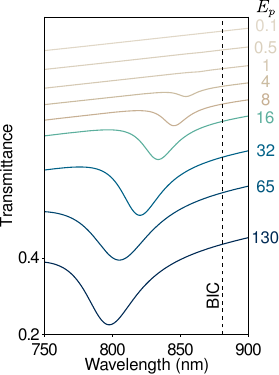}
	\caption{Finite-difference time-domain (FDTD) method-calculated transmittance spectra as a function of the pump pulse energy ($E_p$) per arm (shown on the right). Permittivity parameters used in calculations are the same as stated in the caption of Fig. 2 of the main manuscript.}
  	\label{figs_PvsQ_Fig2c}
\end{figure}

\newpage

\section{Optical response versus the polarization of one of the pump beams}

\begin{figure}[h] 
	\centering
	\includegraphics[scale=0.98]{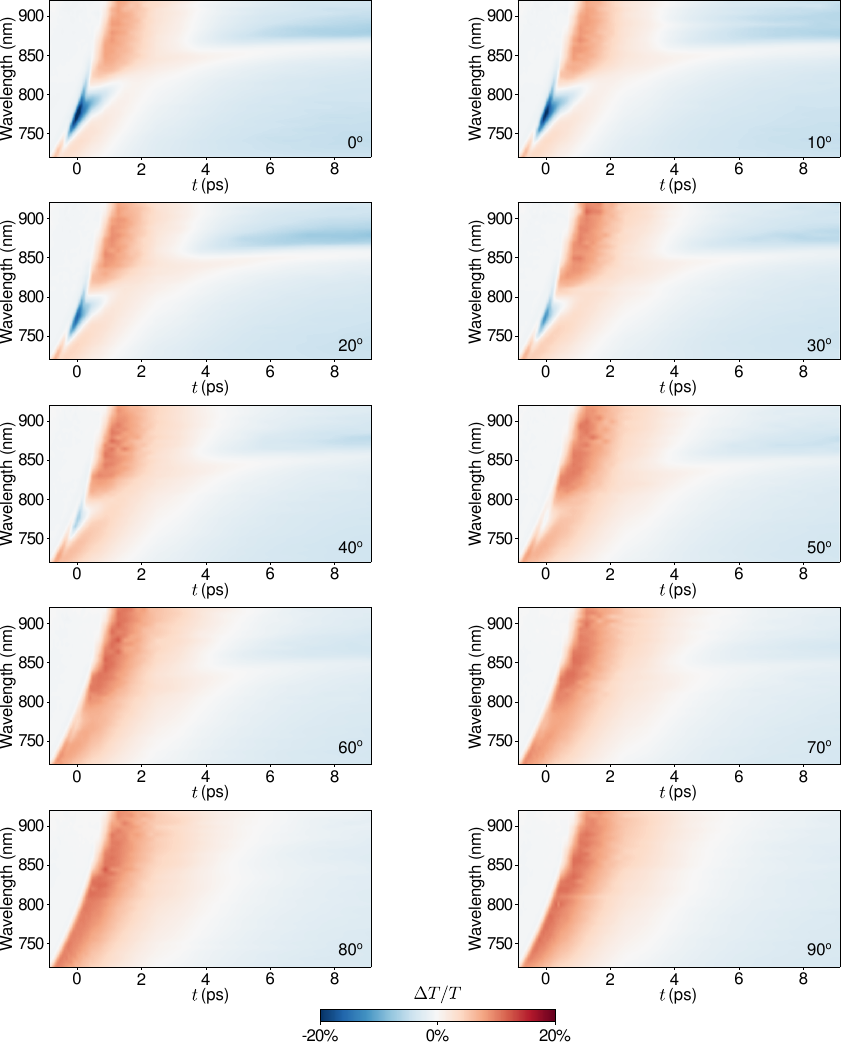}
	\caption{Effect on the $\varepsilon$-\textit{q}BICs of the relative polarization (shown in the insets) between the pump beams, controlled via a half-wave plate positioned in one of the pump arms (as shown in Fig. 1c of the main manuscript). As the relative polarization is increased towards an orthogonal configuration, the optically-induced metasurface generation is hindered, and the mode excitation is less effective. Measurements performed at $\lambda_{pump} =$ 420 nm and 130 nJ pulse energy per arm.}
  	\label{figs2}
\end{figure}

\newpage

\begin{figure}[h] 
	\centering
	\includegraphics[scale=1]{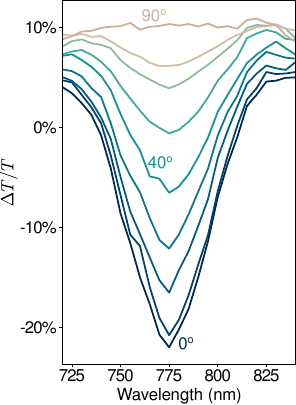}
	\caption{Temporal cross-sections at $t =$ 0 ps of modulated transmittance as a function of the relative polarization between the pump beams (selected angles shown along curves as inset).}
  	\label{figs2_cs}
\end{figure}

\newpage

\section{Temporal evolution of $\varepsilon$-\textit{q}BICs as a function of the probe polarization}

\begin{figure}[h] 
	\centering
	\includegraphics[scale=1]{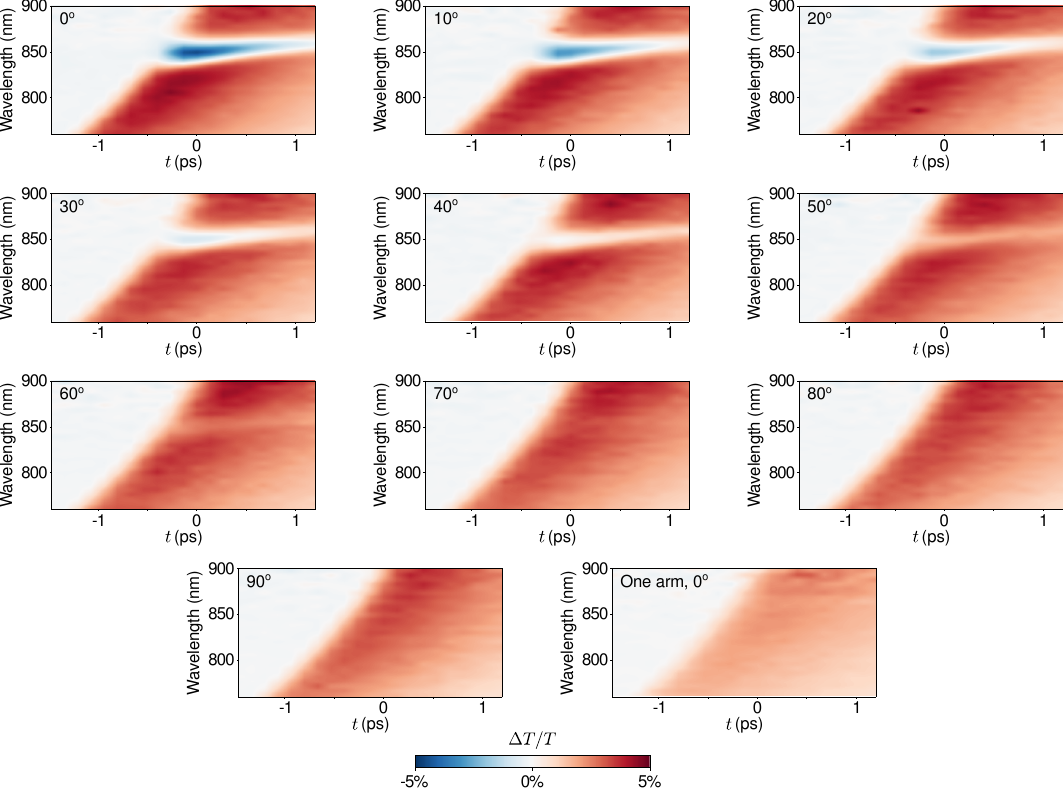}
	\caption{Temporal evolution of $\varepsilon$-\textit{q}BICs as a function of the probe polarization (insets) relative to the $x$-axis of the metasurface (along the ridges). Coupling to the $\varepsilon$-\textit{q}BIC reduces as the relative polarization angle increases. When only one of the pump arms illuminates the thin film (a condition in which no metasurface is formed) no resonance is observed, even at a 0$^{\circ}$ relative polarization. Measurements performed at pump wavelength $\lambda_{pump} =$ 420 nm and 16 nJ pulse energy per arm.}
  	\label{figs_probe_polarization}
\end{figure}

\newpage

\begin{figure}[h] 
	\centering
	\includegraphics[scale=1]{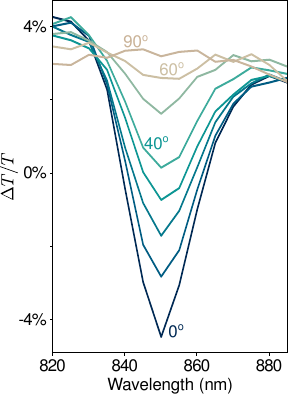}
	\caption{Temporal cross-sections at $t =$ 0 ps of differential transmittance versus probe polarization (inset) relative to the $x$-axis of the metasurface (along the ridges, see Fig. 1 of the main manuscript). Coupling to the $\varepsilon$-\textit{q}BIC reduces as the relative polarization angle increases.}
  	\label{figs_probe_polarization_CS}
\end{figure}

\newpage

\section{Third-harmonic generation for different pump beams configurations}

\begin{figure}[h] 
	\centering
	\includegraphics[scale=1]{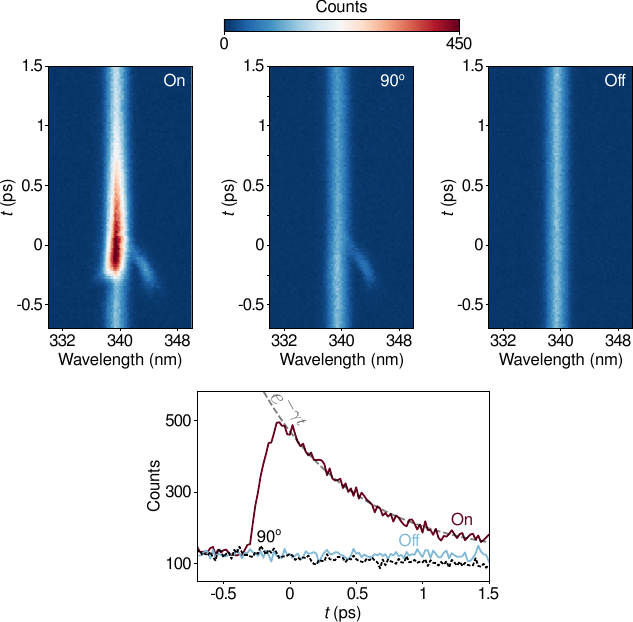}
	\caption{(top) Third harmonic generation (THG) signal versus delay line position (in ps) for different pump configurations (insets). $\lambda_{fundamental} =$ 1020nm. THG enhancement is observed following the formation of the metasurface (0$^{\circ}$, left) and the excitation of an $\varepsilon$-\textit{q}BIC, while a nearly constant signal is obtained for cross-polarized (90$^{\circ}$, center) and in the absence of the pump beams (Off, right). Spot at $\approx$ 343nm in the left and central panels corresponds to the four-wave mixing signal (FWM, $\nu_{fwm} = \nu_{515nm}+\nu_{515nm}-\nu_{1020nm}$), a process due to the overlap of the ultrafast pump and probe beams. (bottom) Time trace of the THG signal at 339 nm for the corresponding configurations. THG enhancement decays exponentially (dashed grey line, $1/e \approx$ 0.72 ps), along with the \textit{quasi}-symmetry-protected resonance excited at low pump powers, not being a process limited by the Gaussian overlap of ultrafast beams. For cross-polarized pump beams (90$^{\circ}$, center), a slight decrease in the THG signal is observed, attributed to increased losses in the film induced by the absorption of the pump beams. Pump pulse energy \mbox{$E_{p,515nm} =$ 75 nJ} per arm, probe pulse energy $E_{p,1020nm} =$ 5 nJ.}
  	\label{figs_thg_Fig3a}
\end{figure}

\newpage

\begin{figure}[h] 
	\centering
	\includegraphics[scale=1]{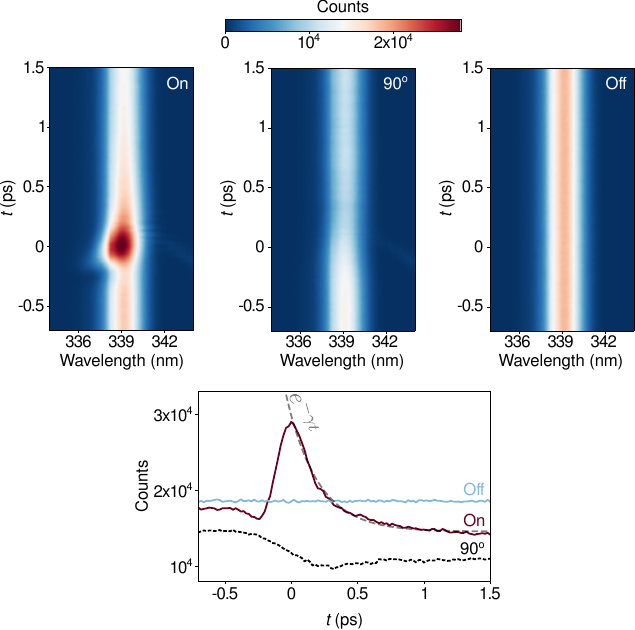}
	\caption{Third harmonic generation (THG) signal at higher probe pulse energy versus delay line position (in ps) for different pump configurations (insets). $\lambda_{fundamental} =$ 1020nm. 
(bottom) Corresponding THG signal as a function of time at \mbox{$\lambda_{THG} =$ 339 nm.}
A similar trend observed at low powers is obtained at high powers, with a faster exponential decay of the THG enhancement process (grey dashed line, $1/e \approx$ 0.24 ps). \mbox{$E_{p,515nm} =$ 50 nJ} per arm, $E_{p,1020nm} =$ 9 nJ.}
  	\label{figs_thg_Fig3d}
\end{figure}

\newpage

\section{Third-harmonic generation versus probe beam polarization}

\begin{figure}[h] 
	\centering
	\includegraphics[scale=0.93]{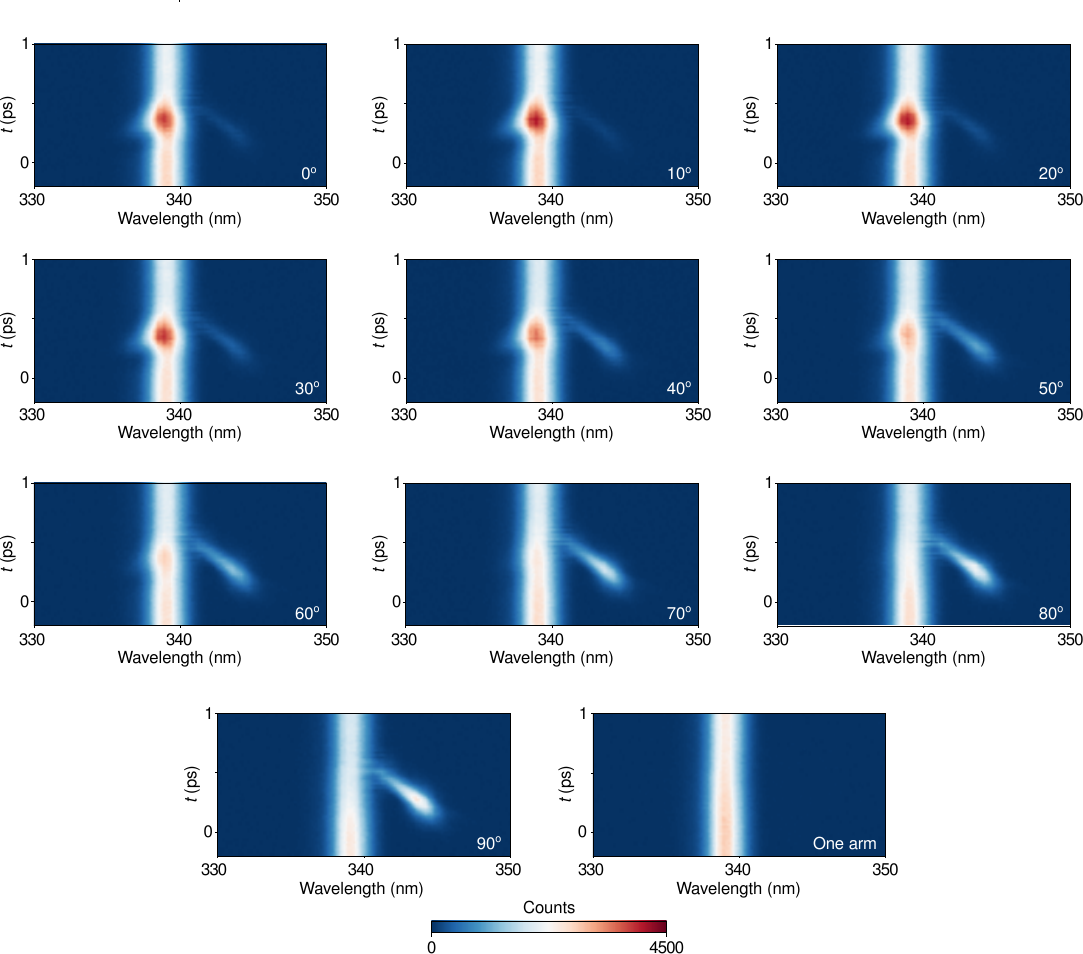}
	\caption{THG and FWM signals vs delay line position (in ps) as a function of the relative polarization between the probe and the pump beams (inset). $\lambda_{fundamental} =$ 1020nm. The efficiency of THG (centered at $\approx$ 339 nm) decreases as the relative polarization angle between the probe and the pump beams increases. Red spot along the THG vertical strip corresponds to the excitation of an $\varepsilon$-\textit{q}BIC and the enhancement of the THG signal. The efficiency of FWM ($\nu_{fwm} = \nu_{515nm}+\nu_{515nm}-\nu_{1020nm}$, at $\approx$ 343 nm), on the other hand, increases with increasing relative polarization. When only one pump arm is used, the efficiency of THG increases (relative to the 90$^{\circ}$ case, due to smaller losses), albeit with no enhancement, and no FWM signal is observed. \mbox{$E_{p,515nm} =$ 50 nJ} per arm, $E_{p,1020nm} =$ 9 nJ.}
  	\label{figs_thg_hwp}
\end{figure}

\newpage

\begin{figure}[h] 
	\centering
	\includegraphics[scale=0.964]{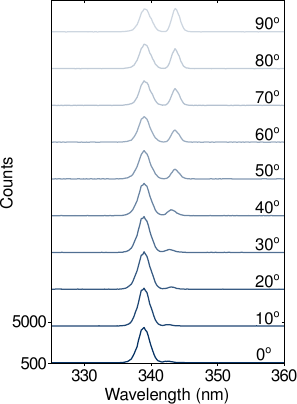}
	\caption{TGH and FWM signals as a function of the relative polarization betweem pump and probe beams. A reduction of the THG signal (peak at $\approx$ 339 nm) occurs for larger relative polarizations while the red-shifted competing FWM process output increases. \mbox{$E_{p,515nm} =$ 50 nJ} per arm, $E_{p,1020nm} =$ 9 nJ.}
  	\label{figs_thg_vs_fwm}
\end{figure}

\newpage

\section{Third-harmonic generation versus probe beam wavelength}

\begin{figure}[h] 
	\centering
	\includegraphics[scale=0.98]{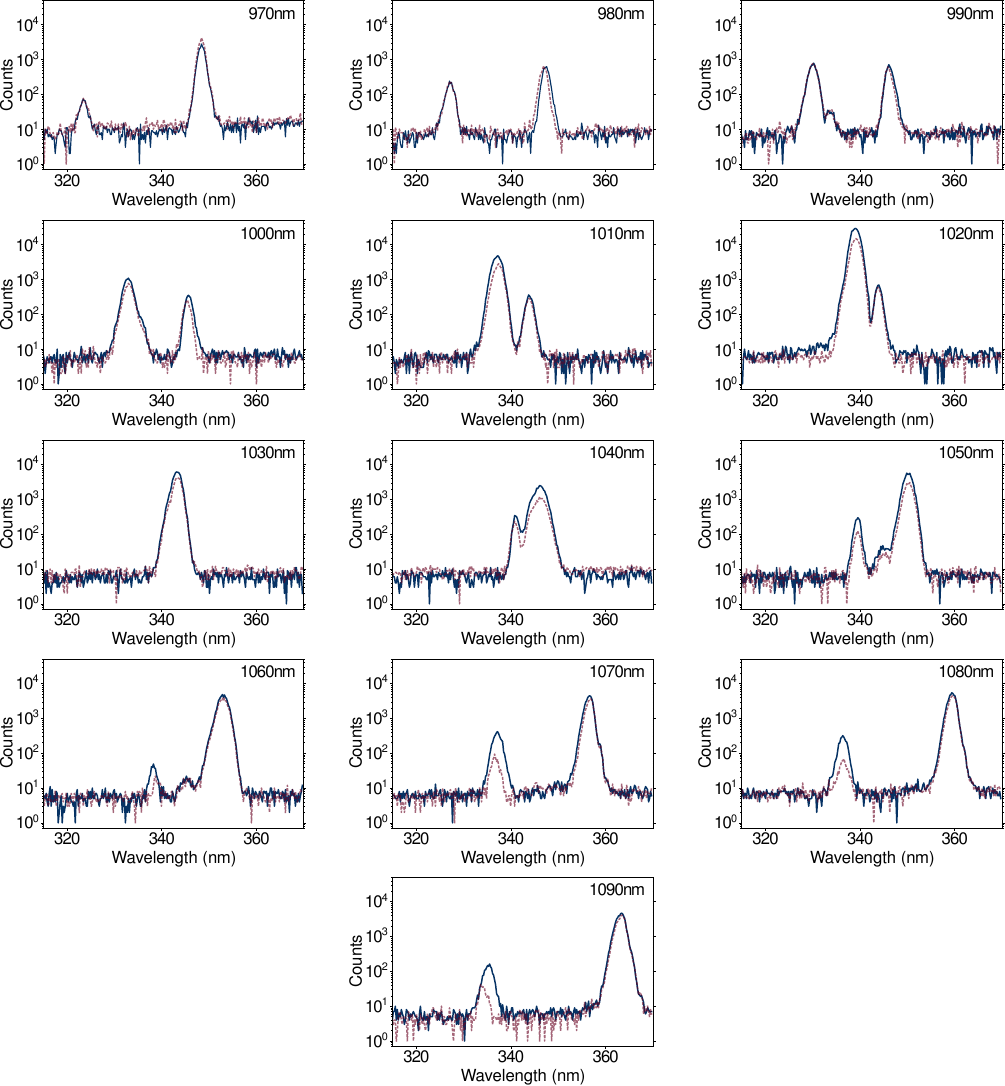}
	\caption{THG and FWM signals vs probe wavelength (inset) for a 0$^{\circ}$ (blue) and a 90$^{\circ}$ (dashed red) relative polarization between pump beams ($\lambda_{pump} =$ 515 nm). THG and FWM red- and blue-shift, respectively (crossing at $\lambda_{probe} =$~1030~nm), for longer probe wavelengths. Note the logarithmic scale in the $y$-axis (Counts).}
  	\label{figs_thg_lambda}
\end{figure}

\newpage

\begin{figure}[h] 
	\centering
	\includegraphics[scale=0.964]{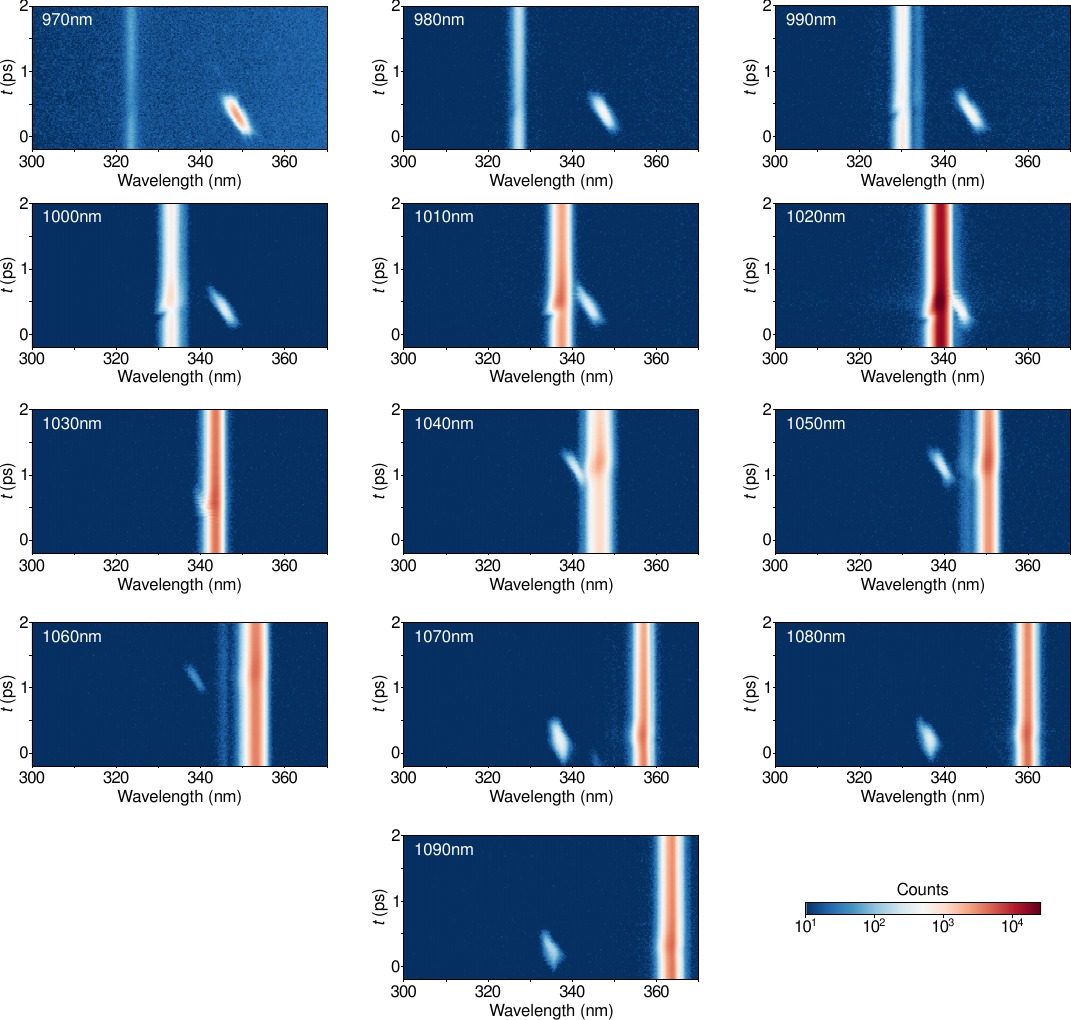}
	\caption{THG and FWM signals vs delay line position (in ps) as a function of the probe wavelength (inset). Here a 0$^{\circ}$ relative polarization between the pump beams is used (thus forming the metasurface). Note the logarithmic scale in the colormap (counts). \mbox{$E_{p,515nm} =$ 50 nJ} per arm, $E_{p,probe} =$ 9 nJ. The simultaneous arrival of beams (red spot along the THG vertical strip) changes in time for $\lambda_{probe} =$ 1040 nm and $\lambda_{probe} =$ 1070 nm as the OPA changes its output beam between signal and idler. The fainter vertical stripes corresponds to both signal and idler beams being present, leading to two closely-spaced THG signals.}
  	\label{figs_thg_lambda_map}
\end{figure}

\newpage

\begin{figure}[h] 
	\centering
	\includegraphics[scale=0.964]{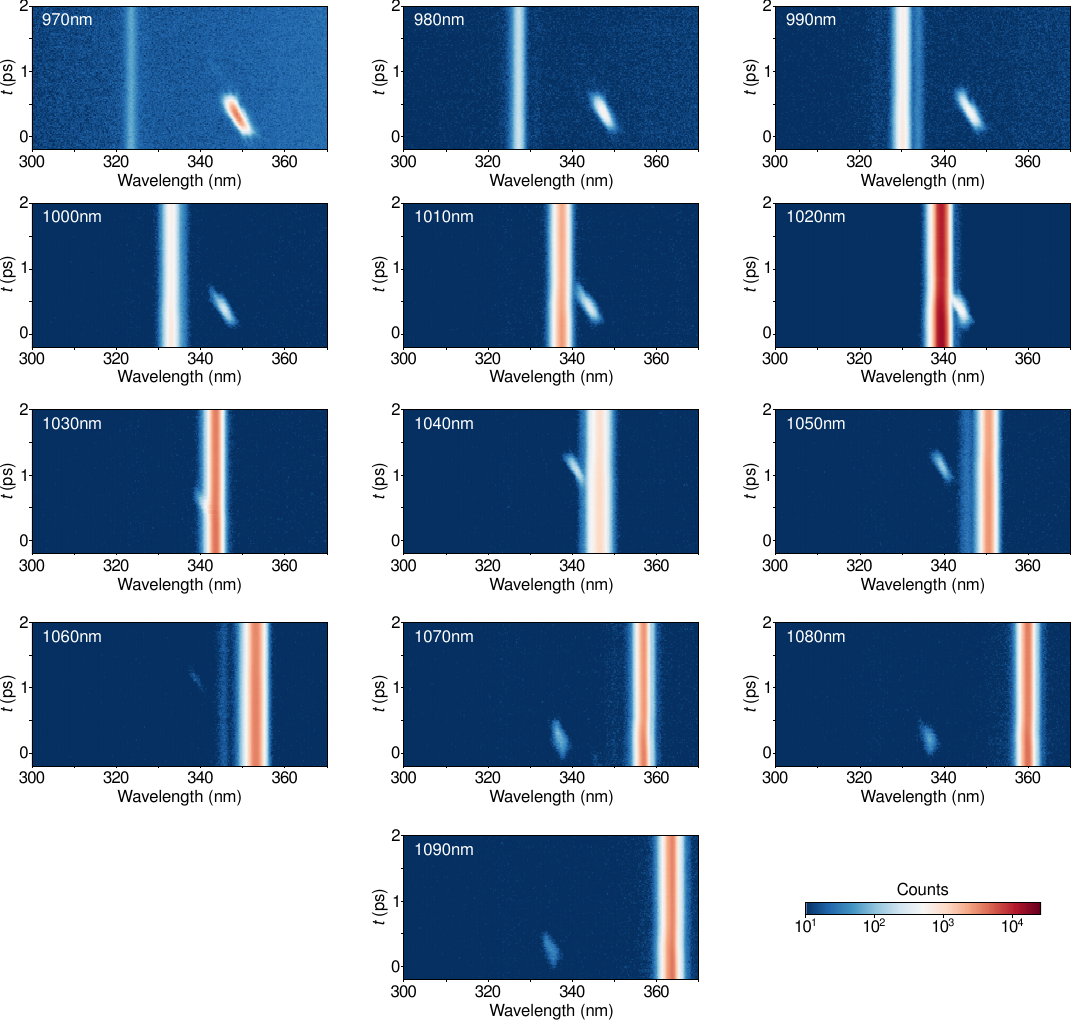}
	\caption{THG and FWM signals vs delay line position (in ps) as a function of the probe wavelength (inset). Here a 90$^{\circ}$ relative polarization between the pump beams is used (absence of the metasurface), reducing the harmonic generation efficiency. Note the logarithmic scale in the colormap (Counts). \mbox{$E_{p,515nm} =$ 50 nJ} per arm, $E_{p,probe} =$ 9 nJ.}
  	\label{figs_thg_lambda_hwp45}
\end{figure}

\newpage

\begin{figure}[h] 
	\centering
	\includegraphics[scale=0.964]{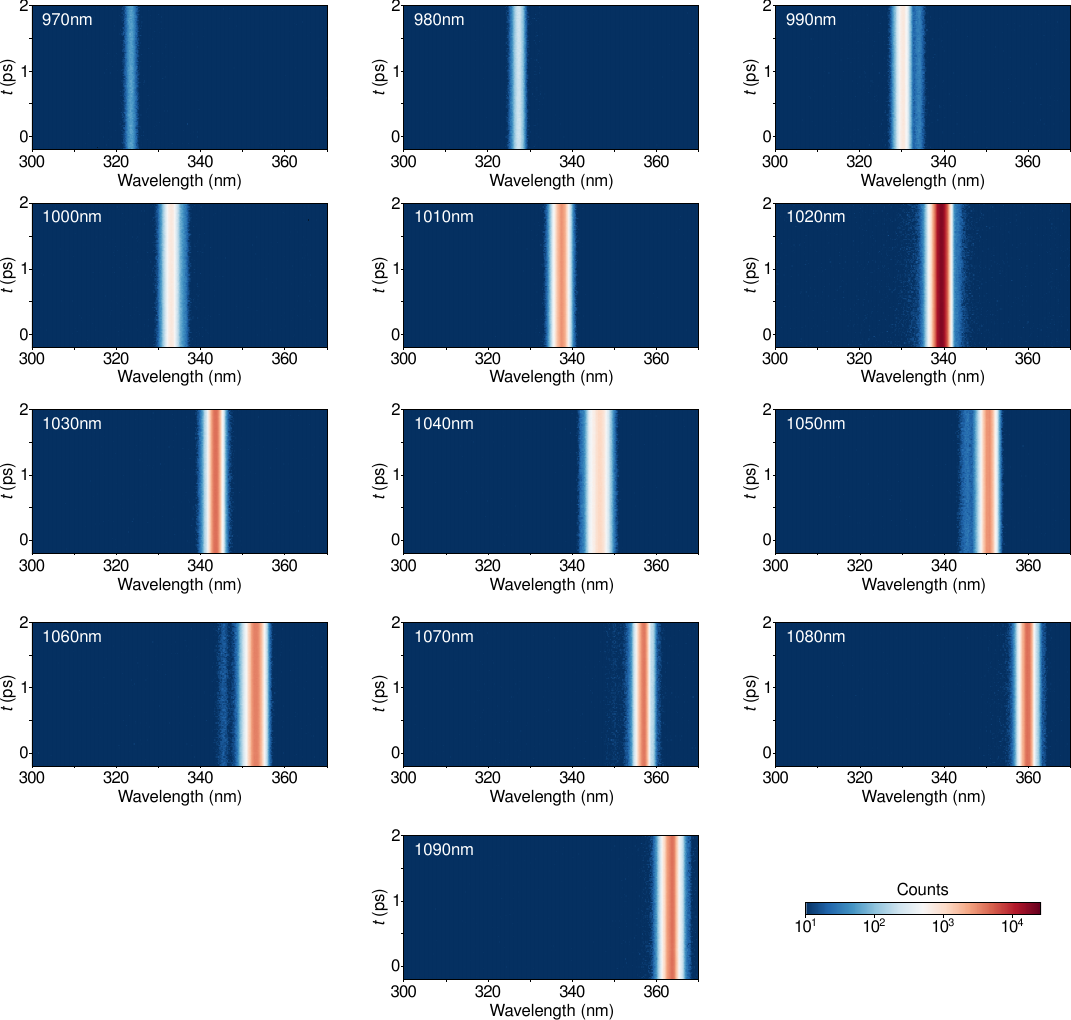}
	\caption{THG and FWM signals vs delay line position (in ps) as a function of the probe wavelength (inset) in the absence of the pump beams. As the 515 nm photons are not present, the FWM signal vanishes. Note the logarithmic scale in the colormap (Counts). $E_{p,probe} =$ 9 nJ.}
  	\label{figs_thg_lambda_pumpoff}
\end{figure}

\newpage


\end{document}